\def \ASTRNACH #1 #2 {{\em Astron. Nach.\/} {\bf #1}, #2}\def \AAP #1 #2 {{\em Astron. Astrophys.\/} {\bf #1}, #2}\def \AAL #1 #2 {{\em Astron. Astrophys. Lett.\/} {\bf #1}, L#2}\def \AAR #1 #2 {{\em Astron. Astrophys. Rev.\/} {\bf #1}, #2}\def \AAS #1 #2 #3  {{\em Astron. Astrophys. Suppl. Ser.\/} {\bf #1}  #2 #3}   % modificata io \def \AJ #1 #2 {{\em Astron. J.\/} {\bf #1}, #2}\def \ANNREV #1 #2 {{\em Ann. Rev. Astron. Astrophys.\/} {\bf #1}, #2}\def \APJ #1 #2 #3 {{\em Astophys. Journal\/} {\bf #1}  #2  #3}   % modificata io \def \APJL #1 #2 {{\em Astrophys.. J. Lett.\/} {\bf #1}, L#2}\def \APJS #1 #2 {{\em Astrophys. J. Suppl.\/} {\bf #1}, #2}\def \APSS #1 #2 {{\em Astrophys. Space Sci.\/} {\bf #1}, #2}\def \ASR #1 #2 {{\em Adv. Space Res.\/} {\bf #1}, #2}\def \BAIC #1 #2 {{\em Bull. Astron. Inst. Czechosl.\/} {\bf #1}, #2}\def \JSQRT #1 #2 {{\em J. Quant. Spectrosc. Radiat. Transfer\/} {\bf #1},#2}\def \MN #1 #2 {{\em Mon. Not. R. Astr. Soc.\/} {\bf #1}, #2}\def \MEM #1 #2 {{\em Mem. R. Astr. Soc.\/} {\bf #1}, #2}\def \PLR #1 #2 {{\em Phys. Lett. Rev.\/} {\bf #1}, #2}\def \PASJ #1 #2 {{\em Publ. Astron. Soc. Japan\/} {\bf #1}, #2}\def \PASP #1 #2 {{\em Publ. Astr. Soc. Pacific\/} {\bf #1}, #2}\def \NAT #1 #2 {{\em Nature\/} {\bf #1}, #2}

\documentclass[10pt]{iopart}
\usepackage[dvips]{graphicx}
\usepackage{epsfig}
\usepackage{amssymb}
\usepackage{amsfonts}

\newcommand{\ds}{{\sffamily DarkSUSY}}

\def\ir{I \kern-0.35em R}
\def\IN{I \kern-0.35em N}
\def\FF{I \kern-0.35em F}
\def\ZZ{Z \kern-0.55em Z}
\def\ic{I \kern-0.65em C}

\begin{document}
 \title[Uncertainties of Cosmic Ray Spectra and Detectability of Antiproton ...]{Uncertainties of Cosmic Ray Spectra and Detectability of Antiproton mSUGRA  Contributions With PAMELA}

\author{A.M. Lionetto, A. Morselli, V. Zdravkovi\'c}
\address{INFN, Sezione di Roma II, via della Ricerca Scientifica 1,Roma, Italy and Dipartimento di Fisica, Universit\`a di Roma "Tor Vergata", via della Ricerca Scientifica 1, Roma, Italy}

\begin{abstract}
We studied the variation of $e^+$ and $\bar p$ top of the atmosphere spectra due to the parameters uncertainties of the Milky Way geometry, propagation models and cross sections. We used the B/C data and Galprop code for the propagation analysis. We also derived the uncertainty bands for subFe/Fe ratio, H, He, antiproton to proton ratio and the positron charge fraction. Finally, we considered a neutralino induced component in the antiproton flux in the mSUGRA framework. PAMELA expectations for  positrons and antiprotons are calculated. 
We studied in details the possibility to disentangle an eventual supersymmetric component in the antiproton spectra in a clumpy halo scenario. We computed the minimal and the maximal values of the clumpiness factors needed to disentangle the signal from the background without violating present data.
The main result of this work is that, assuming as the background the best fit for the DC model, PAMELA will be able to disentangle an eventual supersymmetric signal even for small clumpiness factors.
Lower values of the background  will expand the allowed mSUGRA parameter space but for higher clumpiness factors. Higher values of the background will reduce the allowed mSUGRA parameter space
but for smaller clumpiness factors.

\end{abstract}


\eads{\mailto{andrea.lionetto@roma2.infn.it}, \mailto{aldo.morselli@roma2.infn.it}, \mailto{Vladimir.Zdravkovic@roma2.infn.it}}
\maketitle

%==========================================================
\section{Introduction}
%==========================================================
The scope of this article is two-folded: the first part is dedicated to the analysis of the propagation uncertainties of cosmic rays in the Milky Way; in the second part we studied the detection possibilities of an eventual neutralino induced component in the antiproton spectra in the framework of minimal supergravity (mSUGRA).

In order to disentangle an exotic contribution from the standard one in the antiproton spectra first we have to test the accuracy of the standard calculation.

%To be able to distinguish the part of the antiproton spectra due to the Standard Model (SM) production and the part of it due to some other mechanism or to give some bounds for some of the unknown parameters in the framework of the anyone of the new physical theories, first we should know the accuracy of standard calculations.
%In the case of antiprotons and supersymmetric candidate for dark matter we try to consider both of these problems. For some other species of cosmic rays we treated just the uncertainty of the standard production.

In Section 2 we give a brief description of standard mechanisms for the propagation of cosmic rays and we present the results about the uncertainties in the standard propagation model due to the unknown geometrical and hydromagnetodynamical parameters of the Galaxy, uncertainties of the measurements and parametrizations of nuclear cross sections and uncertainty of the helium to hydrogen ratio in the Galaxy.

In Section 3 we describe creation and propagation of a neutralino induced component.

In Section 4 we study the possibility to detect secondary components in the positron and antiproton fluxes by the upcoming PAMELA satellite experiment.

In Section 5 we present the results about the possibility to disentangle an eventual signal component in the antiproton spectra. We did this analysis in a clumpy halo scenario. Finally we found the minimal and the maximal values of the clumpiness factors needed to disentangle the signal from the background without violating present antiproton data. We also give some examples of total fluxes in comparison with the experimental data.

Section 6 contains the results and the conclusions.

%==========================================================
\section{Propagation of  Cosmic Rays in the Milky Way and Its Uncertainties}
%==========================================================
The most complete equation for the propagation of cosmic rays that includes all the known physical processes is

\begin{eqnarray}
\frac{\partial \psi ({\mathbf r},p,t) }{\partial t} 
&=& q({\mathbf r}, p)+ \nabla \cdot ( D_{xx}\nabla\psi - {\mathbf V_c} \psi )
+ \frac{d}{dp}\, p^2 D_{pp} \frac{d}{dp}\, \frac{1}{p^2}\, \psi \nonumber\\
&-& \frac{\partial}{\partial p} \left[\dot{p} \psi
- \frac{p}{3} \, (\nabla \cdot {\mathbf V_c} )\psi\right]
- \frac{1}{\tau_f}\psi - \frac{1}{\tau_r}\psi\, ,
\label{EQprop}
\end{eqnarray}

where $\psi ({\mathbf r},p,t)$ is the total phase space density. We will shortly review here the main features of the physical processes described by this equation implemented in the Galprop code~\cite{SM134, SM2}.

The second term describes isotropic diffusion, defined by the coefficient that depends on the rigidity (momentum per unit of charge, $\rho = p/Z$)

\begin{equation}
D_{xx} = \beta D_0(\rho/\rho_0)^{\delta},
\label{EQdiffcoef}
\end{equation}
inspired by the Kolmogorov spectrum ($\delta = 1/3$) of the weak magnetohydrodynamic turbulence. In \cite{SH} was first shown that the Kolmogorov spectrum best reproduces the sharp peak in B/C data. In some models we used a break in the index $\delta$ at some reference rigidity $\rho_0$: $\delta_1 = 0$ below $\rho_0$ while $\delta_2 > 0$ above $\rho_0$.

The convection velocity field ${\mathbf V_c}$, that corresponds to the Galactic wind, has a cylindrical symmetry.  Its z-component is the only one different from zero. It increases linearly with the distance $z$ from the Galactic plane. This is in agreement with magnetohydrodynamical models \cite{Z}. In the Galactic plane there should be no discontinuity in the convection velocity field and so we considered only $V_{c}(z=0) = 0$.

Reacceleration is determined by the diffusion coefficient in the momentum space $D_{pp}$. $D_{pp}$ is a function of the corresponding configuration space diffusion coefficient $D_{xx}$ and of the Alfven velocity $V_{A}$ in the framework of quasi-linear MHD theory \cite{F, SP, B}

\begin{equation}
D_{pp}(D_{xx}, V_{A}) = {4 p^2 {V_A}^2 \over 3 \delta (4-\delta^2)(4-\delta) w}\ ,
\end{equation}
where $w$ characterizes the level of turbulence, and it is equal to the ratio of MHD wave energy density to magnetic field energy density. It is assumed $w = 1$, but the only relevant quantity is $V_A^2 / w$.

The unknown values of parameters such as  the Alfven velocity,  the convection velocity gradient in Milky Way and the height of the galactic halo can be constrained by the B/C data.  With the sets of the constrained parameters one can find all the possible spectra for the others cosmic rays. 
This procedure was already used in \cite{FN1, FN2, FN3} for another propagation code.

Injected spectra of all primary nuclei are power laws

\begin{equation}
dq(p)/dp \propto p^{-\gamma},
\label{EQprimspec}
\end{equation}
where $\gamma$ is the injection index. In principle its value can vary with species.
It can be shown that the power law approximation as well as a small break in the injection indexes $\gamma$ is allowed in the framework of diffusive shock acceleration models~\cite{BO, E, SMx}.

Source term $q({\mathbf r}, p)$ for secondaries contains cross sections for their production from progenitors on H and He targets

\begin{equation}
q(\vec r, p) = \beta c\,
\psi_p (\vec r, p)[\sigma^{ps}_H (p) n_H (\vec r)+ \sigma^{ps}_{He}(p)
n_{He}(\vec r)],
\end{equation}
where $\sigma^{ps}_H (p)$ and $\sigma^{ps}_{He} (p)$ are the production cross sections for the secondary from the progenitor on H and He targets, $\psi_p$ is the progenitor density, and $n_H$, $n_{He}$ are the interstellar hydrogen and helium number densities.

The last two terms in equation (\ref{EQprop}) are loss terms with characteristic times $\tau_f$ and $\tau_r$ related respectively to the fragmentation and radioactive decay.
The heliospheric modulation of the local interstellar spectra in the vicinity of the Earth and in the heliosphere hole has to be taken into account in order to obtain the realistic cosmic rays spectra in locations where they are/will be measured (balloon-born or satellite-borne experiments).
We made use of a widely used and tested model in which the transport equation is solved in the force field approximation \cite{GA, P}. That equation describes diffusion processes in the heliosphere and includes effects of heliospheric magnetic field and solar wind. In this case, solar modulation is a function of just a single parameter that describes the strength of the modulation. All the dynamical processes are simulated simply changing the interstellar spectra during the propagation inside the heliosphere:

\begin{equation}
\label{EQpga}
\frac{ \Phi^{toa} (E^{toa})}{\Phi^{is}(E^{is})} =  (\frac{p^{toa}}{p^{is}})^{2} ,
\end{equation}

\begin{equation}
E^{is} -  E^{toa} = |Ze| \phi ,
\end{equation}
where $E$ and $p$ are energies and momenta of the interstellar and of top of the atmosphere fluxes. The solar modulation is entirely determined by the $\phi$ parameter.

%%==========================================================
%\section{Propagation of the Background Component of Cosmic Rays in the Milky Way and Its Uncertainties}
%%==========================================================

In Galprop the model of the Galaxy is three dimensional with cylindrical symmetry; the coordinates are $(R,z,p)$, where $R$ is Galactocentric radius, $z$ is the distance from the Galactic plane, and $p$ is the total particle momentum. The distance from the Sun to the Galactic centre is taken to be 8.5 Kpc. The propagation region is bounded fixing $R_{max} \equiv R=30$ Kpc and  $z_{max} \equiv z$ beyond which free escape is assumed.  The distribution of cosmic rays sources is chosen to reproduce (after propagation) the cosmic rays distribution determined by the analysis of EGRET gamma-ray data done in \cite{SMx}. The code first computes propagation of primaries, giving the primary distribution as a function of ($R, z, p$). Then the secondary source function is obtained from the gas density and cross sections. Finally, the secondary propagation is computed.

\begin{figure}
\centering
\includegraphics[width=10cm]{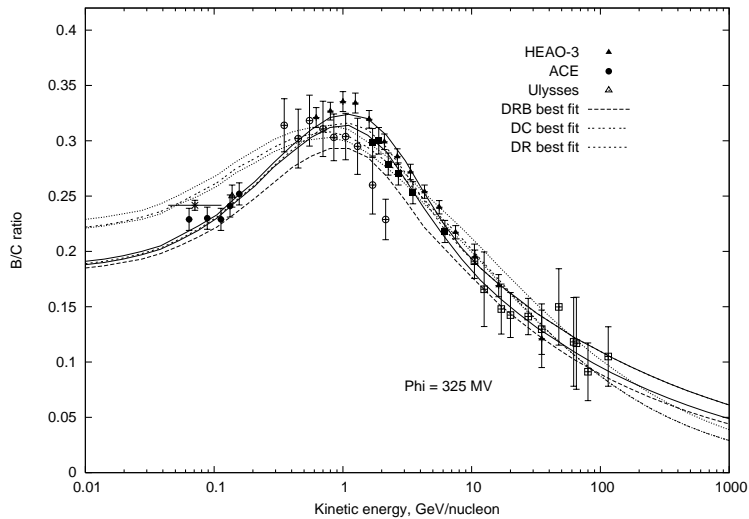}
\caption{Propagation parameters uncertainty for B/C ratio: for the DR model it is given with solid lines around the best fit (dashed line), while for the DC model it is given with dotted lines around the best fit (dashed line). For DRB model we give the best fit (dashed line). For the complete list of the experimental data see \cite{expBC}.}
\label{figBC}
\end{figure}

Secondary to primary CR ratios are the most sensitive quantities on variation of the propagation parameters. This can be verified numerically. Primary to primary and secondary to secondary ratios are not very sensitive to changes in the propagation parameters because they have similar propagation mechanisms.  The most accurately measured secondary to primary ratio is boron to carbon ratio (B/C, see~\cite{expBC}). Boron is secondary while, one of its progenitors, carbon is primary. The B/C data are used also because they have relatively well known cross sections. To estimate the quality of the data fit we used the standard $\chi^2$ test
\begin{equation}
\chi^2 = \frac{1}{N-1} \sum_{n} \frac{1}{(\sigma^{B/C}_n)^2} (\Phi^{B/C}_{n,exp} - \Phi^{B/C}_{n,teo})^2,
\end{equation}
where $\sigma^{B/C}$ are statistical errors for $N=46$ experimental points and $\Phi^{B/C}_{exp}$ are measured and $\Phi^{B/C}_{teo}$ are predicted values of the ratio. 

We analyzed only  the two extreme cases of the propagation models: either without convection or without reacceleration. We did not succeed in obtaining satisfactory models with all the physical processes switched on. A possible explanation is that one of the processes is really subdominant in the Galaxy in such a way that the data require either reacceleration or convection to be well fitted. 
 So in the first case we included diffusive and reacceleration effects (see equation (\ref{EQprop}) and we considered two sub-cases: (DR) without a break in the index of primary injection spectra and (DRB) with a break at the rigidity $\rho^{\gamma}$.
The second case (DC) contains diffusion and convection terms from the same equation (\ref{EQprop}). In this model there are two breaks: one in the index of the primary injection spectra and the other in the spectra of the diffusion coefficient $D_{xx}$. The first break means that at some rigidity $\rho^{\gamma}$ the index $\gamma$ suffers a discontinuity (see equation (\ref{EQprimspec}) and ref. \cite{E} for details), while the second one means that at some rigidity $\rho_{0}$ the diffusion index $\delta$ suffers a discontinuity (see equation (\ref{EQdiffcoef}) and discussion in ref. \cite{SM2}). 
 
\begin{table}
\caption{Allowed values for diffusion and reacceleration model propagation parameters.}
\begin{center}
\renewcommand{\arraystretch}{1}
\setlength\tabcolsep{10pt}
\begin{tabular}{||c||c|c|c|c|c|c|c||}
\hline\hline
par./val. & $z[Kpc]$ & $D_{0}[cm^2 s^{-1}]$ & $\delta$ & $\gamma$ & $v_A[Kms^{-1}]$ \\ \hline\hline
minimal & 3.0 & 5.2 $10^{28}$ & 0.25 & 2.35 & 22 \\ \hline
best fit & 4.0 & 5.8 $10^{28}$ & 0.29 & 2.47 & 26 \\ \hline
maximal & 5.0 & 6.7 $10^{28}$ & 0.36 & 2.52 & 35 \\ \hline\hline
\end{tabular}
\end{center}
\label{tab1}
\end{table}

\begin{figure}
\centering
\includegraphics[width=10cm]{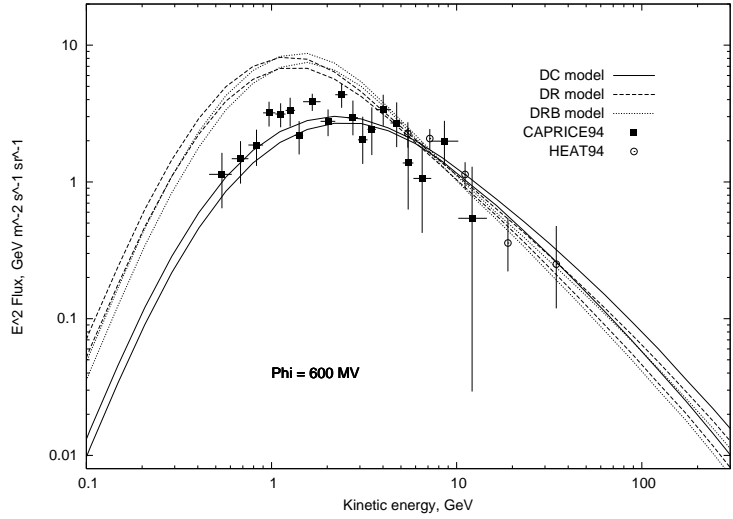}
\caption{Upper and lower bounds of positron spectra due to the uncertainties of the propagation parameters are represented with solid lines for the DC model, with dashed lines for the DR model and with dotted lines for the DRB model. The DR and DRB model uncertainties are very similar, but there is a slight improvement of the fit in the low energy part of the spectra in the case of DRB model. On the other hand, around the maximum the DRB model overestimates the data slightly more than the DR model. Experimental data are taken from~\cite{expe}.}
\label{fige+par}
\end{figure}

\begin{figure}
\centering
\includegraphics[width=10cm]{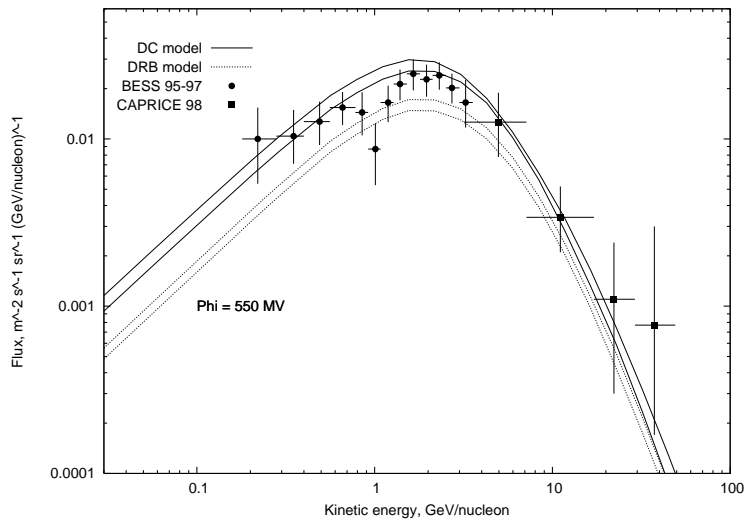}
\caption{Upper and lower bounds of the antiproton spectra due to the uncertainties of the propagation parameters for the DC model (solid lines) and DRB model (dotted lines). The propagation parameters uncertainties for the DR model are almost the same as those of the DRB model (see the discussion later in the text), so we are not presenting them here to avoid the confusion. Experimental data are taken from~\cite{exppbar}.}
\label{figpbarpar}
\end{figure}

For the DR model we chose to vary the following parameters: the height of the Galactic halo $z$, the constant in the diffusion coefficient $D_{0}$ (from equation~(\ref{EQdiffcoef})), the index of the diffusion coefficient $\delta$ (from the same equation), the primary spectra injection index $\gamma$ for all the energies (from equation~(\ref{EQprimspec})) and the Alfven velocity $v_A$ that determines the strength of reacceleration. 
In order to find the allowed range of these parameters we have required a reduced $\chi^2$ less than 2 for the fit of the B/C experimental data~\cite{expBC}.
In figure~\ref{figBC} are presented the enveloping curves of all the good fits with solid lines around the best fit line for the same model, that is represented with dashed line. We took the experimental data with relatively small solar modulation parameter $\phi$ between 325 MV and 600 MV, where the force field approximation is better justified than for higher modulation parameters. The allowed ranges of propagation parameters of this model are given in the table~\ref{tab1}. Using this result we found the enveloping curves of all the positron and antiproton spectra. In this way we determined the upper and the lower bounds of the uncertainty bands for positron and antiproton spectra due to the propagation parameters uncertainty (see figure~\ref{fige+par} for the positron spectra uncertainty and figure~\ref{figpbarpar} for the antiproton spectra uncertainty). The relative uncertainty depends on the energy range of the spectra. For positrons the relative uncertainty varies from 30\% under 1 GeV to 15\% around 10 GeV, increasing again after 10~GeV (figure~\ref{fige+par}) while for antiprotons it varies from about 10\% up to about 15\% (figure~\ref{figpbarpar}).

For a better visualization some figures that contain positron and antiproton data contain just the part of the experimental data chosen to cover in a convenient way all the energy range reached in experiments. The data are taken from references~\cite{exppbar, expe}.

\begin{table}
\caption{Allowed values for the propagation parameters for diffusion convection model.}
\begin{center}
\renewcommand{\arraystretch}{1}
\setlength\tabcolsep{10pt}
\begin{tabular}{||c||c|c|c|c|c|c||}
\hline\hline
par./val.& $z[Kpc]$ & $D_{0}[\frac{cm^2}{s}]$ & $\delta_2$ & $\frac{dV_C}{dz}[\frac{Km}{skpc}]$ & $\gamma_1$ & $\gamma_2$ \\ \hline\hline
minimal & 3.0 & 2.3 $10^{28}$ & 0.48 & 5.0 & 2.42 & 2.14 \\ \hline
best fit & 4.0 & 2.5 $10^{28}$ & 0.55 & 6.0 & 2.48 & 2.20 \\ \hline
maximal & 5.0 & 2.7 $10^{28}$ & 0.62 & 7.0 & 2.50 & 2.22 \\ \hline\hline
\end{tabular}
\end{center}
\label{tab2}
\end{table}

In the case of the DC model we chose to vary the following parameters: $D_0$, the diffusion index $\delta_1$ below the reference rigidity $\rho_0=4$~GV and $\delta_2$ above it (see equation~(\ref{EQdiffcoef})), the halo size $z$, the convection velocity $V_c$ (see equation~(\ref{EQprop})) and the injection index for primary nuclei $\gamma_1$ below the reference rigidity $\rho^{\gamma}_0=20$~GV and $\gamma_2$ above it (see equation~(\ref{EQprimspec})). Enveloping curves of the B/C fits for a reduced $\chi^2$ values less than 2.8 are given in figure~\ref{figBC}. Positive variations around $\delta_1=0$ gave unsatisfactory fit. In order to take the smallest possible break of this index we decided not to take negative $\delta_1$ values. Allowed values for the propagation parameters can be found in table~\ref{tab2}. Again, we used them to derive the uncertainties of positron and antiproton spectra. Relative uncertainty for positrons vary between 20\% above the maximum and 30\% below it~(figure \ref{fige+par}) while for antiprotons is about 20\% around 20~MeV, 17\% around the maximum and 25\% around 20~GeV (figure~\ref{figpbarpar}).

\begin{table}
\caption{Allowed values for the propagation parameters for diffusion reacceleration model with break in the injection spectra of primary nuclei.}
\begin{center}
\renewcommand{\arraystretch}{1}
\setlength\tabcolsep{10pt}
\begin{tabular}{||c||c|c|c|c|c|c|c||}
\hline\hline
par./val. & $z[Kpc]$ & $D_{0}[\frac{cm^2}{s}]$ & $\delta$ & $\gamma_1$ & $\gamma_2$ & $v_{A}[\frac{Km}{s}]$ \\ \hline\hline
minimal & 3.5 & 5.9 $10^{28}$ & 0.28 & 1.88 & 2.36 & 25 \\ \hline
best fit & 4.0 & 6.1 $10^{28}$ & 0.34 & 1.92 & 2.42 & 32 \\ \hline
maximal & 4.5 & 6.3 $10^{28}$ & 0.36 & 2.02 & 2.50 & 33 \\ \hline\hline
\end{tabular}
\end{center}
\label{tab3}
\end{table}

\begin{figure}
\centering
\includegraphics[width=11cm]{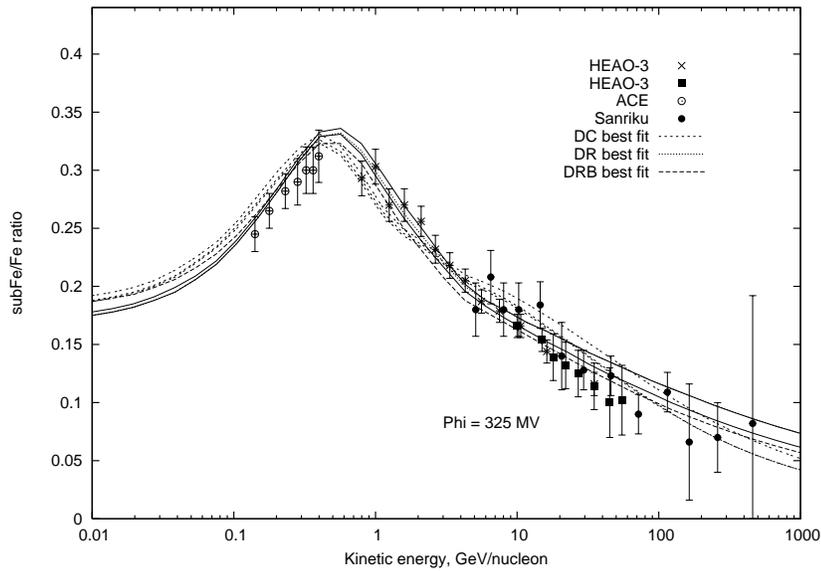}
\caption{Ratio (Sc+Ti+V)/Fe that corresponds to the propagation parameters that give the best fits of B/C data for the DC model given with dashed line. It is inside the corresponding uncertainty band also given with dashed lines. The ratio for  the DR model is given with dotted line and it is inside the uncertainty given with solid lines, while for DRB model is given with larger-step dashed line without the uncertainty band around. Experimental data are taken from \cite{expsubFe/Fe}.}
\label{figFe}
\end{figure}

\begin{figure}
\centering
\includegraphics[width=11cm]{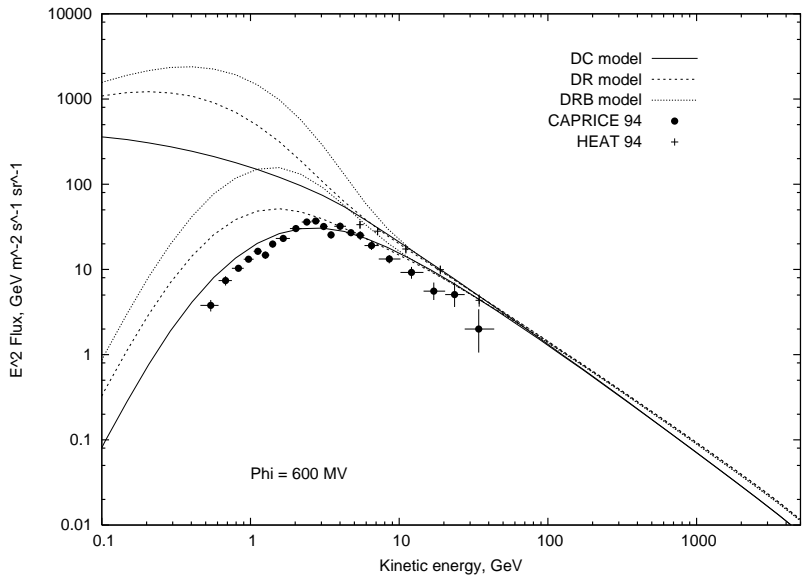}
\caption{Top of the atmosphere spectra of electrons that correspond to the parameters of the best B/C fit are the lower curves: for the DC model they are given with solid line, for the DR model with dashed line and for DRB model with dotted line. The local interstellar spectra are the upper curves; the three models are represented with the same types of lines. Experimental data are taken from \cite{expe}.}
\label{fige-}
\end{figure}

\begin{figure}
\centering
\includegraphics[width=11cm]{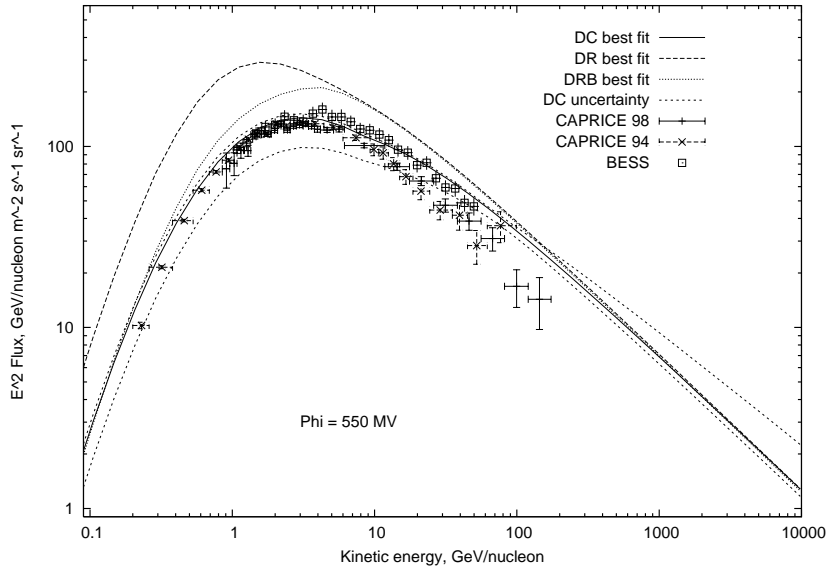}
\caption{Upper and lower bounds of helium spectra due to the uncertainties of the propagation parameters for the DC model are represented with dashed lines. Spectra that correspond to the parameters of the best B/C fit are given for the DC model with solid line, for the DR model with dashed line and for the DRB model with dotted line. Experimental data are taken from~\cite{exppHe}.}
\label{figHe}
\end{figure}

\begin{figure}
\centering
\includegraphics[width=11cm]{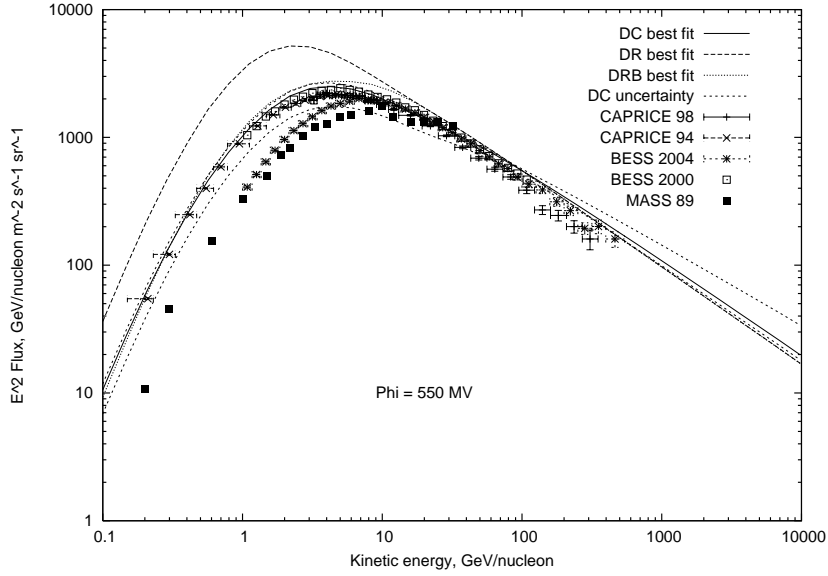}
\caption{Upper and lower bounds of proton spectra due to the uncertainties of the propagation parameters for the DC model are represented with dashed lines. Spectra that correspond to the parameters of the best B/C fit are given for the DC model with solid line, for the DR model with dashed line and for the DRB model with dotted line.  Experimental data are taken from~\cite{exppHe}.}
\label{figp}
\end{figure}

We also found the spectra that correspond to the parameters of the best fit of B/C data for subFe/Fe ratio (see figure \ref{figFe}), protons, helium and electrons as well as the corresponding propagation parameters uncertainties. 
  The solar activity (around a minimum in 1998 and around a maximum in 2002) is the cause of the discrepancies in the data below 10 GeV for protons.  
 For DC model the fits are good, while DR overestimates protons (figure \ref{figp}), helium (figure \ref{figHe}) and electrons (figure \ref{fige-}).

In order to improve those fits, we considered the DR model with a break in the injection index for the primary nuclei spectra with a rigidity of 10~GV~\cite{SM2, E}. We determined the allowed values of the propagation parameters (table~\ref{tab3}) demanding the same reduced $\chi^2$=2 as for the DR model (see figure \ref{figBC}). The positron and antiproton uncertainties are presented in figures \ref{fige+par} and \ref{figpbarpar}. Even if positrons at low energies and protons and helium in all the energy range gives a better fit (see figure \ref{figp} and figure \ref{figHe}), they remain overestimated. For the computation of B/C ratio, Galprop uses only one principal progenitor and compute weighted cross sections. Introducing the break in the index of the primary injection spectra in the DR model give worst electron data fit (see figure \ref{fige-}) than in the case without the break. The antiproton spectra remain unchanged, still significantly and systematically underestimated in all the energy range.

We also calculated the variation of the antiproton spectra as a function of the most important antiproton production cross sections.
Antiprotons are created in the interactions of the primary cosmic rays (protons and other nuclei)  with the interstellar gas. Dominant processes are interactions of high energy primary protons with hydrogen, $p + p \rightarrow p+ p + p + \bar p$. Parametrization of this cross section is given in ~\cite{TN}. Other cross sections, those of the primary protons with other nuclei, are studied in reference~\cite{GS}. From these, the most important are those that involve helium, and they contribute less than 20\% of the total production of all the antiprotons. All the heavier nuclei together give just a few percents of the total production.

Uncertainties of cross sections influence the antiproton spectra uncertainties: simultaneous settings of all the production cross sections to the maximum/minimum rise/lower the upper/lower propagation parameters uncertainty bounds (already found before). Errors obtained in this way give contributions to the total uncertainties: they vary from 20\% up to 25\% in the case of a DR model and, very similarly, from 20\% up to 24\% for a DC model (depending on the energy range of the spectra). Non-production cross sections, the so called tertiary component, correspond to inelastically scattered secondaries $\bar p + X \rightarrow \bar p + \tilde X$. Those processes bring down the energies of the antiprotons of relatively high energies, flattening like that the spectra. But, even if the uncertainty related to those cross sections is relatively big, this does not give relevant change of the antiproton spectra because tertiary contribution is very small. In fact, it has been implemented in propagation codes just recently.

%%%%%%%%%%%%%%%%%%%%%
\begin{figure}
\centering
\includegraphics[width=12cm]{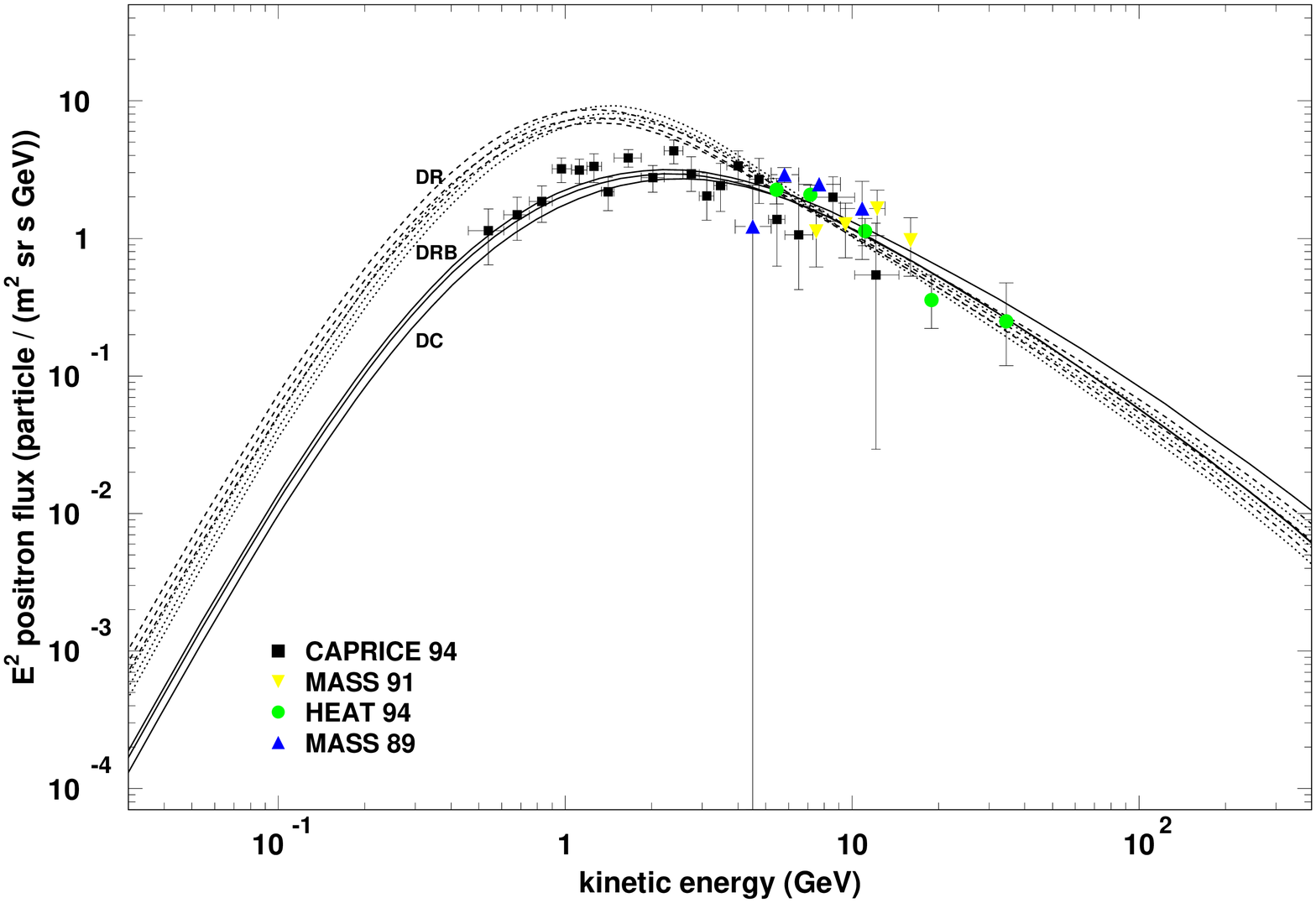}
\caption{Total uncertainties of positron fluxes and spectra that correspond to the parameters of the best B/C fit for the DC model (solid lines around the best fit curve, also solid), DR model (dashed lines around the best fit curve, also dashed) and the DRB model (dotted lines around the best fit curve, also dotted). Experimental data are taken from~\cite{expe}}
\label{fige+toterr}
\end{figure}
%%%%%%%%%%%%%%%%%%%%%
%%%%%%%%%%%%%%%%%%%%%
\begin{figure}
\centering
\includegraphics[width=12cm]{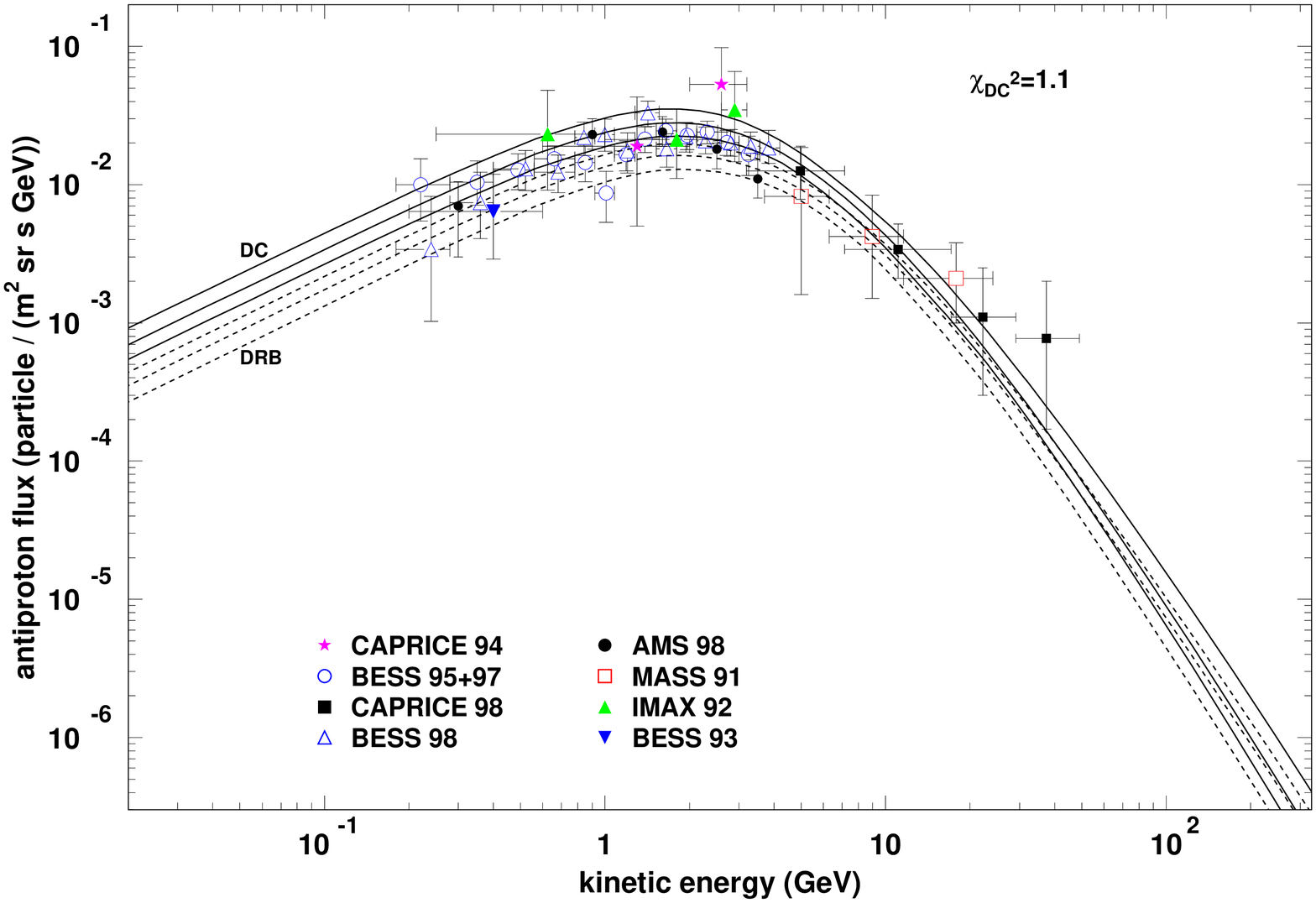}
\caption{Total uncertainties of antiproton fluxes and spectra that correspond to the parameters of the best B/C fit for DC (solid lines around the best fit curve, also solid) and DRB model (dotted lines around the best fit curve, also dotted). Experimental data are taken from \cite{exppbar}}
\label{figpbartoterr}
\end{figure}
%%%%%%%%%%%%%%%%%%%%%

The uncertainty in the measurements of helium to hydrogen ratio brings another component to the total uncertainty of the cosmic rays spectra. By changing the He/H ratio in a reasonable range from 0.08 to 0.11 (see~\cite{Grevesse} and~\cite{Hernandez94}) we obtained a relatively small contribution. For both positron and antiproton flux uncertainty and for all of the models we considered it varies from 3\% to 7\%, depending on the energy.

Total uncertainties of positrons and antiprotons are presented in figure~\ref{fige+toterr} and in figure~\ref{figpbartoterr} respectively. They vary from 35\% up to 55\% for antiprotons and from 20\% up to 40\% for positrons for both the models in the current experimental data energy range.

\begin{figure}
\centering
\includegraphics[width=10cm]{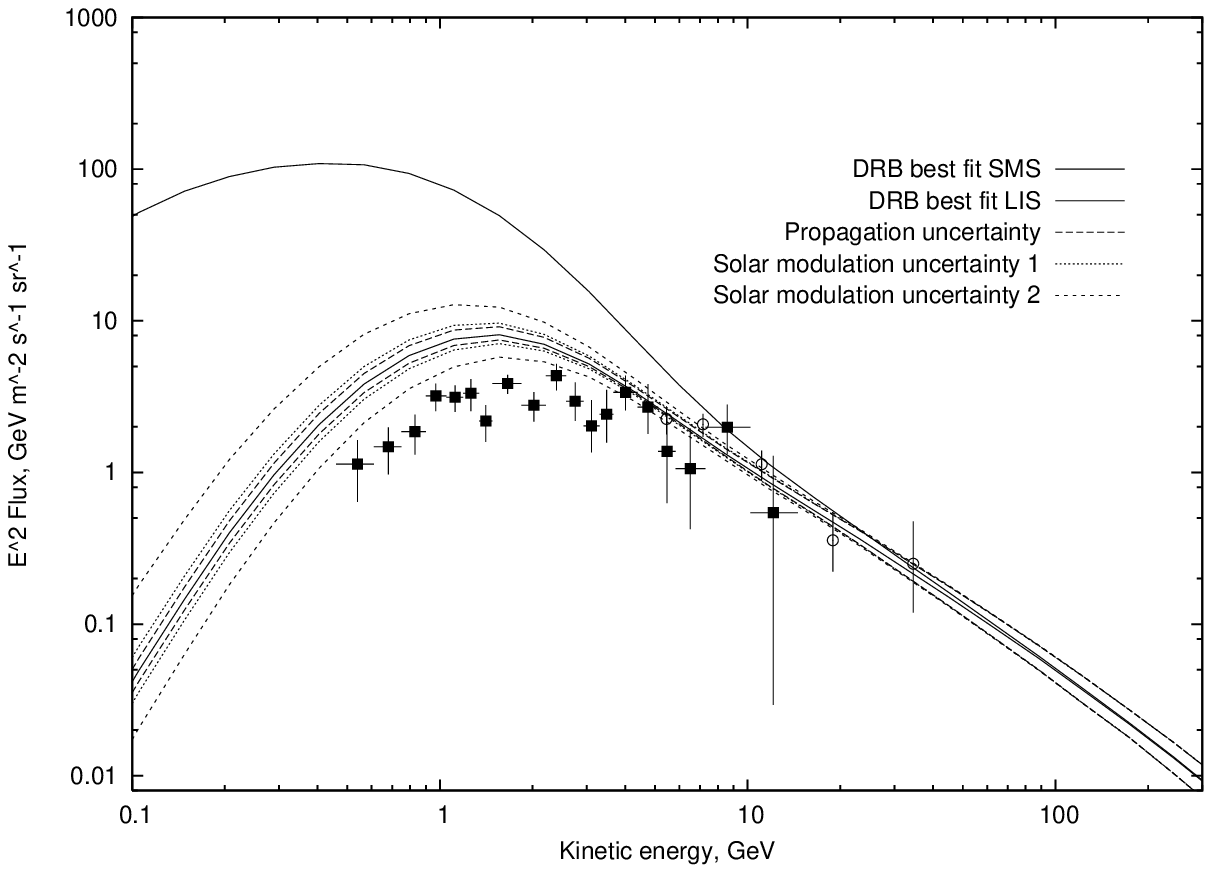}
\caption{Top of the atmosphere (lower solid curve) and local interstellar (upper solid curve) spectra of positrons that correspond to the parameters of the best B/C fit for DRB model. Total uncertainties of positron fluxes are presented with dashed lines around the best fit. Two enlargements of the propagation uncertainties due to two different solar modulation uncertainties are presented with dotted and dashed lines (see the text for details). Experimental data are taken from~\cite{expe}.}
\label{FIGsmee+}
\end{figure}

In plot~\ref{FIGsmee+} we presented two different uncertainties of the positron spectra due to the solar modulation for the same DRB model (that fits better the data than the DR model). The first uncertainty is obtained solar-modulating the lower bound of the propagation uncertainty band using the 10\% bigger $\phi$ than the measured one and the upper bound with 10\% lower modulation parameter. This is a conservative error for the solar modulation parameter. The spectra are still completely above the data. As we used the force field approximation for the solar modulation, we tried also to change the modulation parameter for $\pm$ 50\%. The resulting curves are represented on the plot and are still completely above the data. As we deal with a medium solar modulation ($\phi \approx 600MV$) for positron data (we are not near the solar minimum) the sign-charge drift effect due to the solar magnetic field polarity should be not very strong.

%==========================================================
\section{Component of the Antiproton Spectra Induced by Neutralino Annihilations}
%==========================================================
In this section we take into account the possibility of a neutralino induced component in the $\bar{p}$ flux. Our analysis is performed in the well known mSUGRA framework~\cite{msugra} with the usual gaugino mass universality at the grand unification scale $M_{GUT}$.

In the general framework of the minimal supersymmetric extension of the Standard Model (MSSM), the lightest neutralino is the lightest mass eigenstate obtained from the superposition of four interaction eigenstates, the supersymmetric partners of the neutral gauge bosons (the bino and the wino) and Higgs bosons (two Higgsinos). Its mass, composition and couplings with Standard Model particles and other superpartners are functions of several free parameters one needs to introduce to define such supersymmetric extension. In the mSUGRA model, universality at the grand unification scale is imposed. With this assumption the number of free parameters is limited to five
\[
m_{1/2},\;\;  m_0,\;\;  sign(\mu),\;\;  A_0\;\; \rm{and}\;\; \tan\beta \;,
\]
where $m_0$ is the common scalar mass, $m_{1/2}$ is the common gaugino mass and $A_0$ is the proportionality factor between the supersymmetry breaking trilinear couplings and the Yukawa couplings. $\tan\beta$ denotes the ratio of the VEVs of the two neutral components of the SU(2) Higgs doublet, while the Higgs mixing $\mu$ is determined (up to a sign) by imposing the Electro-Weak Symmetry Breaking (EWSB) conditions at the weak scale. In this context the MSSM can be regarded as an effective low energy theory. The parameters at the weak energy scale are determined by the evolution of those at the unification scale, according to the renormalization group equations (RGEs)~\cite{Cesarini:2003nr}.

For this purpose, we have made use of the ISASUGRA RGE package in the ISAJET 7.64 software~\cite{isasugra}. After fixing the five mSUGRA parameters at the unification scale, we extract from the ISASUGRA output the weak-scale supersymmetric mass spectrum and the relative mixings. Cases in which the lightest neutralino is not the lightest supersymmetric particle or there is no radiative EWSB  are disregarded.

The ISASUGRA output is then used as an input in the \ds\ package \cite{ds}. The latter is exploited to:

\begin{itemize}

\item reject models which violate limits recommended by the Particle Data Group 2002 (PDG)~\cite{pdg02};

\item compute the neutralino relic abundance, with full numerical solution of the density evolution equation including resonances, threshold effects and all possible coannihilation processes~\cite{newcoann}; 

\item compute the neutralino annihilation rate at zero temperature in all kinematically allowed tree-level final states (including fermions, gauge bosons and Higgs bosons);

\item \ds estimates the induced antiproton yield by linking to the results of the simulations performed with the Lund Monte Carlo program Pythia \cite{pythia}.

\end{itemize}

This setup as well as some other similar scenarios were already considered in the context of dark matter detection and of an improvement of the cosmic rays data fits (a list of references includes, for example,~\cite{fengms, NewUllioBott}). The comparison of our results with previous works and other complementary techniques should be transparent.

%==========================================================
\subsection{Clumpy Halo Models}
%==========================================================
 The dependence of the antiproton flux is $\propto \rho^{2}/m_{\chi}^{2}$ (see (\ref{eq:pbarsusyflux}).
 In the case of  a small clump scenario~\cite{Bergstrom:1998jj} for the dark matter halo in our Galaxy
 there is a local density enhancement without increasing the total halo mass and then an increase of the antiproton flux. 

By hypothesis the clump is a spherical symmetric compact object with mass $M_{cl}$ and some density profile $\rho_{cl}\left(\vec{r}_{cl}\right)$. We denote with $f$ the dark matter fraction concentrated in clumps and we introduce the dimensionless parameter $d$
\begin{equation}
d=\frac{1}{\rho_0}\frac{\int d^3 r_{\rm cl}\left[\rho_{\rm
      cl}\left(\vec{r}_{\rm cl}\right)\right]^2}{\int d^3 {
      r_{\rm cl}}\rho_{\rm cl}\left(\vec{r}_{\rm cl}\right)}
\end{equation} 
that gives the overdensity due to a clump with respect to the local halo density $\rho_0=\rho(r_0)$, where $r_0$ is our distance from the Galactic Center (GC). In a smooth halo scenario the total neutralino induced $\bar{p}$ flux calculated for $r=r_o$ is given by~\cite{Bergstrom:1999jc} 
\begin{equation}
  \Phi_{\bar{p}}(r_0,T) \equiv
  (\sigma_{\rm ann}v) \sum_{f}^{}\frac{dN^{f}}{dT}B^{f}
  \left(\frac{\rho_0}{m_{\tilde{\chi}}}\right)^{2}
  \, C_{\rm prop}(T) \;.
\label{eq:pbarsusyflux}
\end{equation}
where $T$ is the $\bar{p}$ kinetic energy, $\sigma_{\rm ann}v$ is the total annihilation cross section times the relative velocity, $m_\chi$ is the neutralino mass, $B^{f}$ and $dN^{f}/dT$, respectively, the branching ratio and the number of $\bar{p}$ produced in each annihilation channel $f$ per unit energy and $C_{\rm prop}(T)$ is a function entirely determined by the propagation model. In the presence of many small clumps the $\bar{p}$ flux is given by
\begin{equation}
\Phi^{\rm clumpy}_{\bar{p}}(r_0,T)=fd\cdot \Phi_{\bar{p}}(r_0,T)
\end{equation}
For the smooth profile we assumed a Navarro, Frenck and White profile (NFW)~\cite{navarro}.

%==========================================================
\subsection{Propagation of the Neutralino Induced Component}
%==========================================================
The primary contribution to the antiproton flux is computed using the public code \ds \cite{Gondolo:2004sc}. We modified the antiproton propagation in order to be consistent with the DC propagation model as implemented in Galprop code. We assumed diffusion coefficient spectra used in Galprop code with our best fit values for the diffusion constants $D_0$ and $\delta$. In \ds, the convection velocity field is constant in the upper and the lower Galactic hemispheres (with opposite signs, and so it suffers unnatural discontinuity in the Galactic plane), while Galprop uses magnetohydrodynamically induced model, in which one component of velocity field along the Galactic latitude (the only one that is different from zero) increases linearly with the Galactic latitude~\cite{Z}. We assumed an averaged convection velocity calculated from the Galactic plane up to the Galactic halo height $z$.

%==========================================================
\section{Detection of the Secondary Components in the Positron and the Antiproton Fluxes by PAMELA}
%==========================================================
In this section we calculate the statistical errors for the PAMELA \cite{AP1, AP2} experiment  for the positron and the antiproton background spectra calculated with Galprop \footnote{The list of the people and the institutions involved in the collaboration together with the on-line status of the project is available at {\sl http://wizard.roma2.infn.it/}.}.

%%%%%%%%%%%%%%%%%%%%%                   GF_f
\begin{figure}[ht]
\centering
\includegraphics[width=10cm]{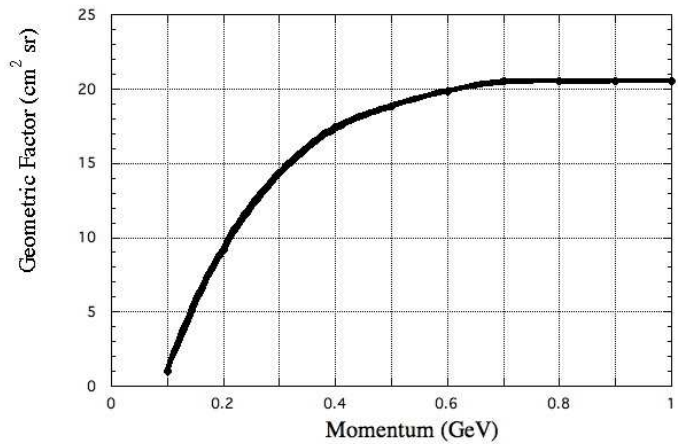}
\caption{Geometric factor of PAMELA}
\label{GF_f}
\end{figure}
%%%%%%%%%%%%%%%%%%%%%

\begin{figure}
\centering
\includegraphics[width=12.5cm]{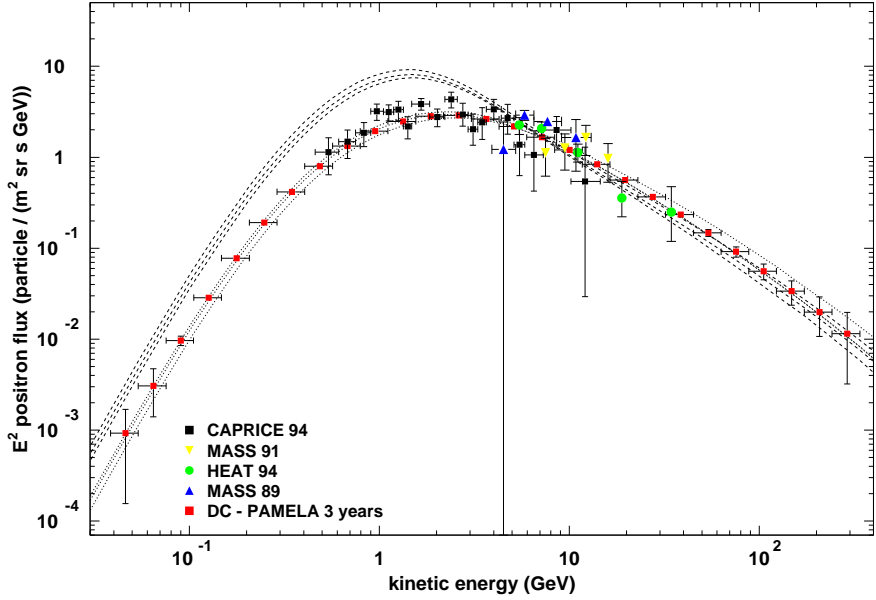}
\caption{Experimental data (from \cite{expe}) confronted with PAMELA's expectations for positrons for DC model background. Total uncertainties of positron fluxes with the spectra that correspond to the parameters of the best B/C fit in the middle are given for a better comparison: for DC model they are represented with dotted lines while for DRB model with dashed lines.}
\label{FIGe+pamDC}
\end{figure}

\begin{figure}
\centering
\includegraphics[width=12.5cm]{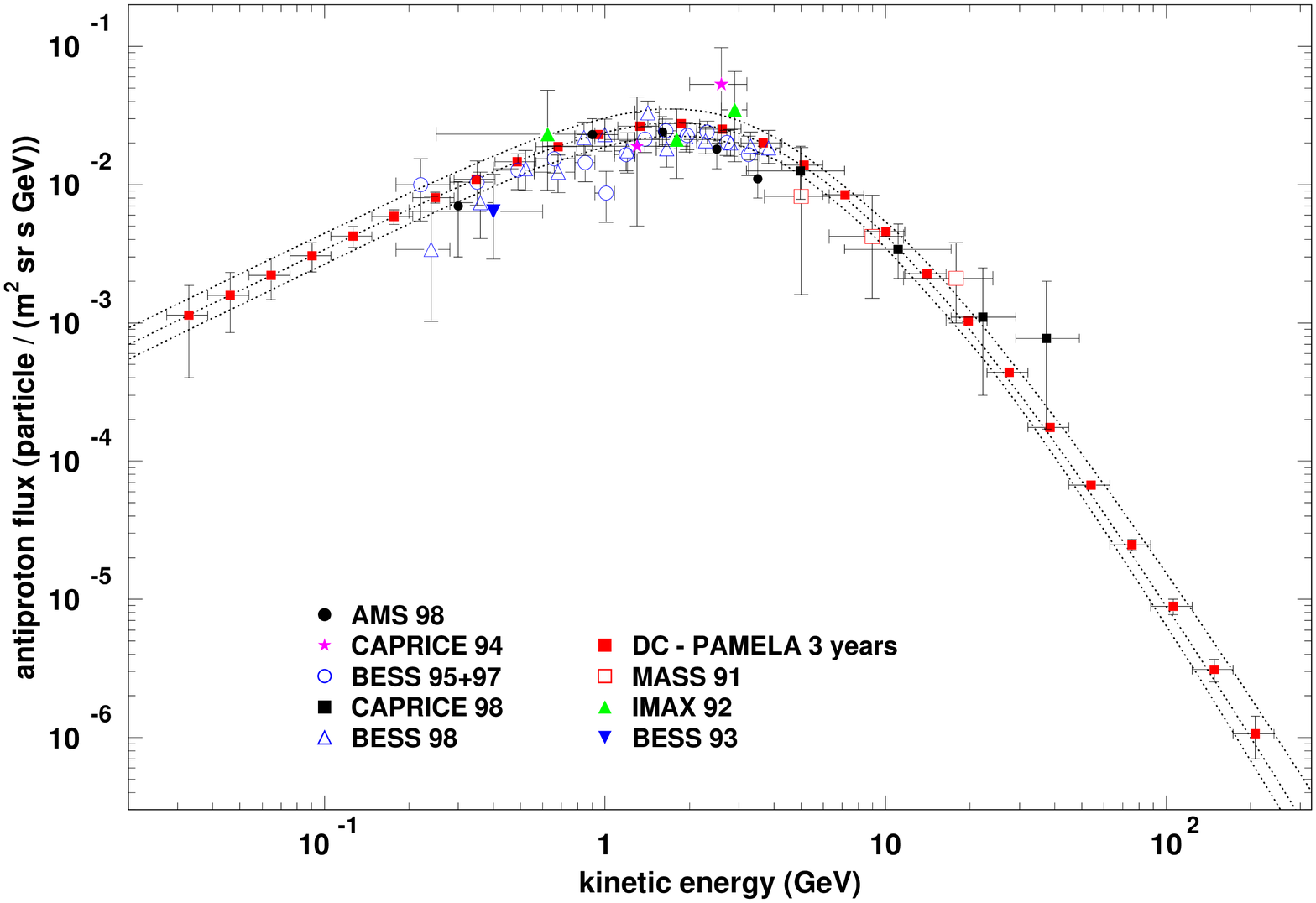}
\caption{Experimental data (from \cite{exppbar}) confronted with PAMELA's expectations for antiprotons for DC model background. Total uncertainty of the DC model antiproton background flux with the spectra that correspond to the parameters of the best B/C fit in the middle are given for a better comparison; they are represented with dotted lines. The reduced $\chi^2$ is 1.1.}
\label{FIGpamDCpbar}
\end{figure}

\begin{figure}
\centering
\includegraphics[width=12.5cm]{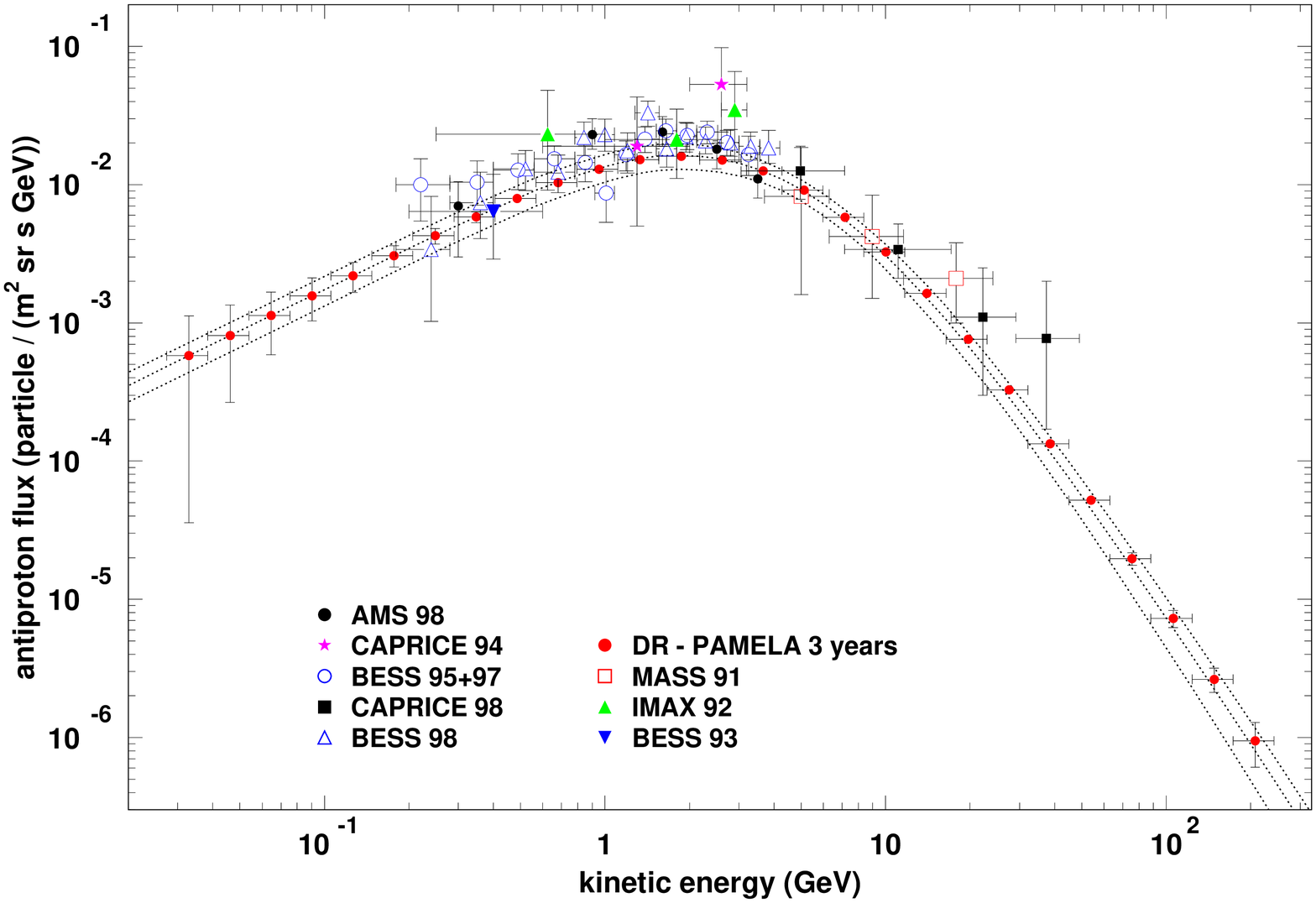}
\caption{Experimental data (from \cite{exppbar}) confronted with PAMELA's expectations for antiprotons for DRB model background. Total uncertainty of the DRB model antiproton background flux with the spectra that correspond to the parameters of the best B/C fit in the middle are given for a better comparison; they are represented with dotted lines.}
\label{FIGpamDRBpbar}
\end{figure}

\begin{figure}
\centering
\includegraphics[width=12.5cm]{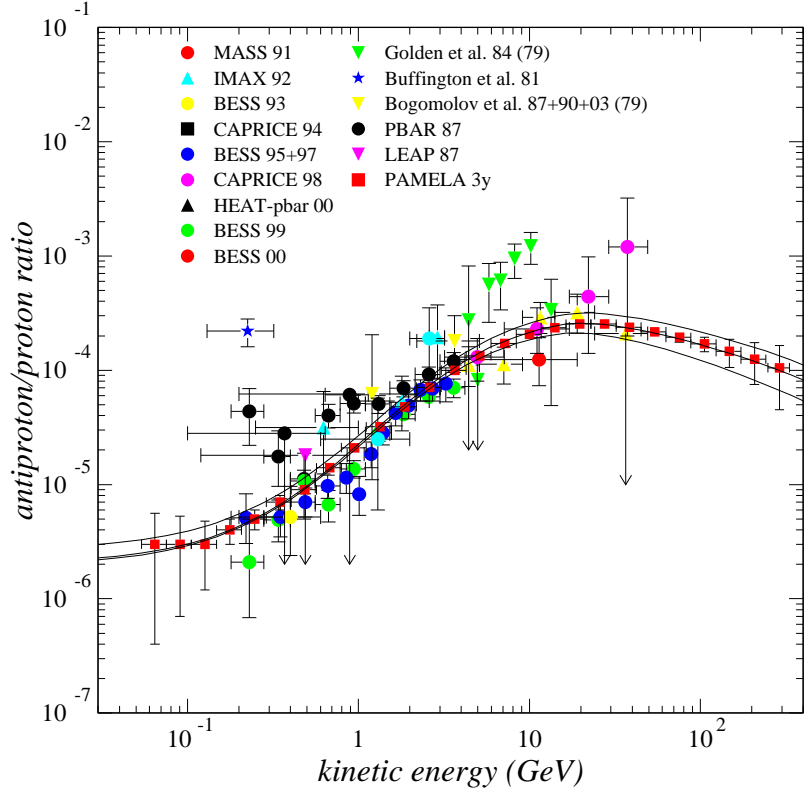}
\caption{Experimental data (from \cite{exppbar} and \cite{exppHe}) confronted with PAMELA's expectations for the antiproton proton ratio for the DC model background. The propagation uncertainty band of the antiproton proton ratio and the curve that corresponds to the parameters of the best B/C fit in the middle are given for a better comparison.}
\label{FIGpbarratio}
\end{figure}

\begin{figure}
\centering
\includegraphics[width=13cm]{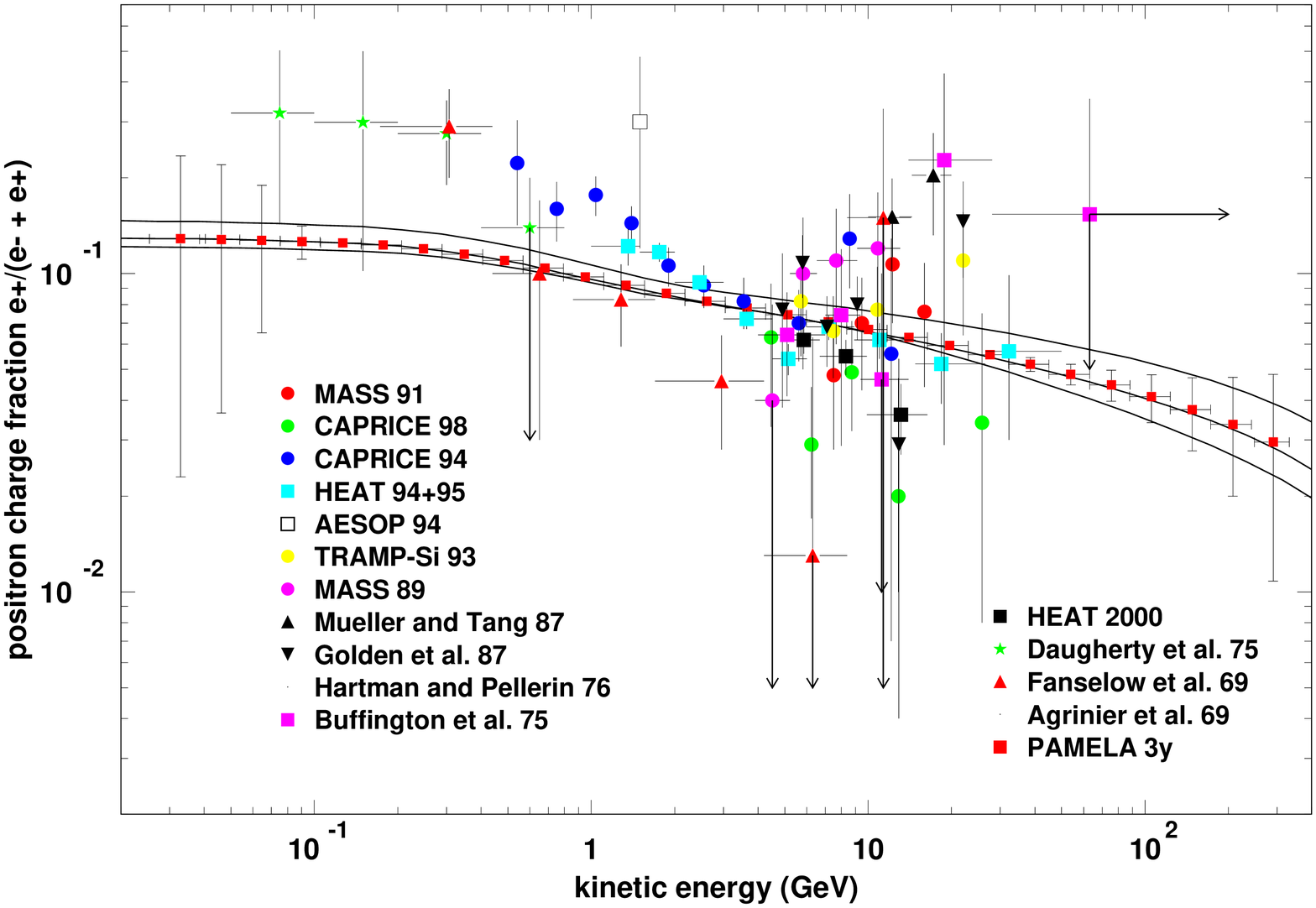}
\caption{Experimental data (from \cite{expe}) confronted with PAMELA's expectations for positron charge fraction for the DC model background. The propagation uncertainty band of the positron charge fraction and the curve that corresponds to the parameters of the best B/C fit in the middle are given for a better comparison.}
\label{FIGe+ratio}
\end{figure}

The calculation is done for a three years mission assuming the geometric factor  given in figure \ref{GF_f} taken from the results of the test measurements on  triggers with the first and the last scintillators  of the apparatus \cite{Papini}. PAMELA's statistical errors for positrons and antiprotons in the case of the DC model are given in figure \ref{FIGe+pamDC} and in figure \ref{FIGpamDCpbar}. We present in figures~\ref{FIGpbarratio} and~\ref{FIGe+ratio} PAMELA expectations together with the propagation uncertainties for the antiproton proton ratio and positron charge fraction, respectively. 
For completeness we also present the expectations for the DRB propagation model for the antiprotons in figure~\ref{FIGpamDRBpbar}.
We do not present PAMELA's expectations for antiprotons in the DR model case because the DR model flux that correspond to the best B/C fit propagation parameters is almost identical with that of DRB model. DR and DRB model background PAMELA predictions for positron fluxes are not given because those fluxes largely overestimate the experimental data at low energies.

%%%%%%%%%%%%%%%%%%%%%                   e+
%%%%%%%%%%%%%%%%%%%%%
%%%%%%%%%%%%%%%%%%%%%                   pbar
%%%%%%%%%%%%%%%%%%%%%
%%%%%%%%%%%%%%%%%%%%%

%%%%%%%%%%%%%%%%%%%%%                   pbar/p
%%%%%%%%%%%%%%%%%%%%%
%%%%%%%%%%%%%%%%%%%%%                   e+/(e+ + e-)
%%%%%%%%%%%%%%%%%%%%%

%==========================================================
\section{The Possibility of Disentanglement of the Neutralino Induced Component in the Antiproton Flux With PAMELA}
%==========================================================
In this section we present the results we found about the minimal values of the clumpiness factors $fd$ needed to disentangle a neutralino induced component in the antiproton flux with PAMELA. We computed this factor as a function of the mSUGRA parameters, fixing $A_{0}$, $\tan \beta$ and ${\rm sign} (\mu) = +1$. In this way the clumpiness factor become a function of $m_{0}$ and $m_{1/2}$ parameters. Similar analysis were already made in the literature (see for example~\cite{NewUllioBott}).

For the discrimination we requested the following conditions:

\begin{enumerate}
\item The total antiproton flux $\phi_{tot} = \phi_{bkg} + \phi_{susy}$ gives a good fit of the experimental data.
\item Difference between $\phi_{tot}$ and the DC model $\phi_{bkg}$ is detectable by PAMELA.
\end{enumerate}
The first condition is satisfied if

\begin{equation}
\chi_{fit}^2 = \frac{1}{N-1} \sum_{n} \frac{( \Phi_{n}^{exp} - \Phi^{tot}_{n} )^2}{(\sigma _{n}^{exp}) ^ 2}\le \chi_{fit, 0}^2
\label{EQc1}
\end{equation} 
where $\chi_{fit, 0}^2 = 1.7$, for $N=40$ experimental points. The second condition is satisfied if

\begin{equation}
\chi_{discr}^2 = \frac{1}{M -1} \sum_{m} \frac{( \Phi_{m}^{bkg} - \Phi^{tot}_{m} )^2}{(\sigma _{m}^{P, bkg}) ^ 2}\ge \chi_{discr, 0}^2
\label{EQc2}
\end{equation}
where $\chi_{discr, 0}^2 = 1.8$, for $M=29$ points and where $\sigma _{m}^{P, bkg}$ are the PAMELA statistical errors associated to the background flux (those presented in figure \ref{FIGpamDCpbar}).

The reduced $\chi_{fit}^2$ for the background flux alone is $\chi_{fit}^2=1.1$. This means that the background already gives a good fit of the experimental data.

%==========================================================
\subsection{Results}
\begin{figure}
\centering
\includegraphics[scale=0.25]{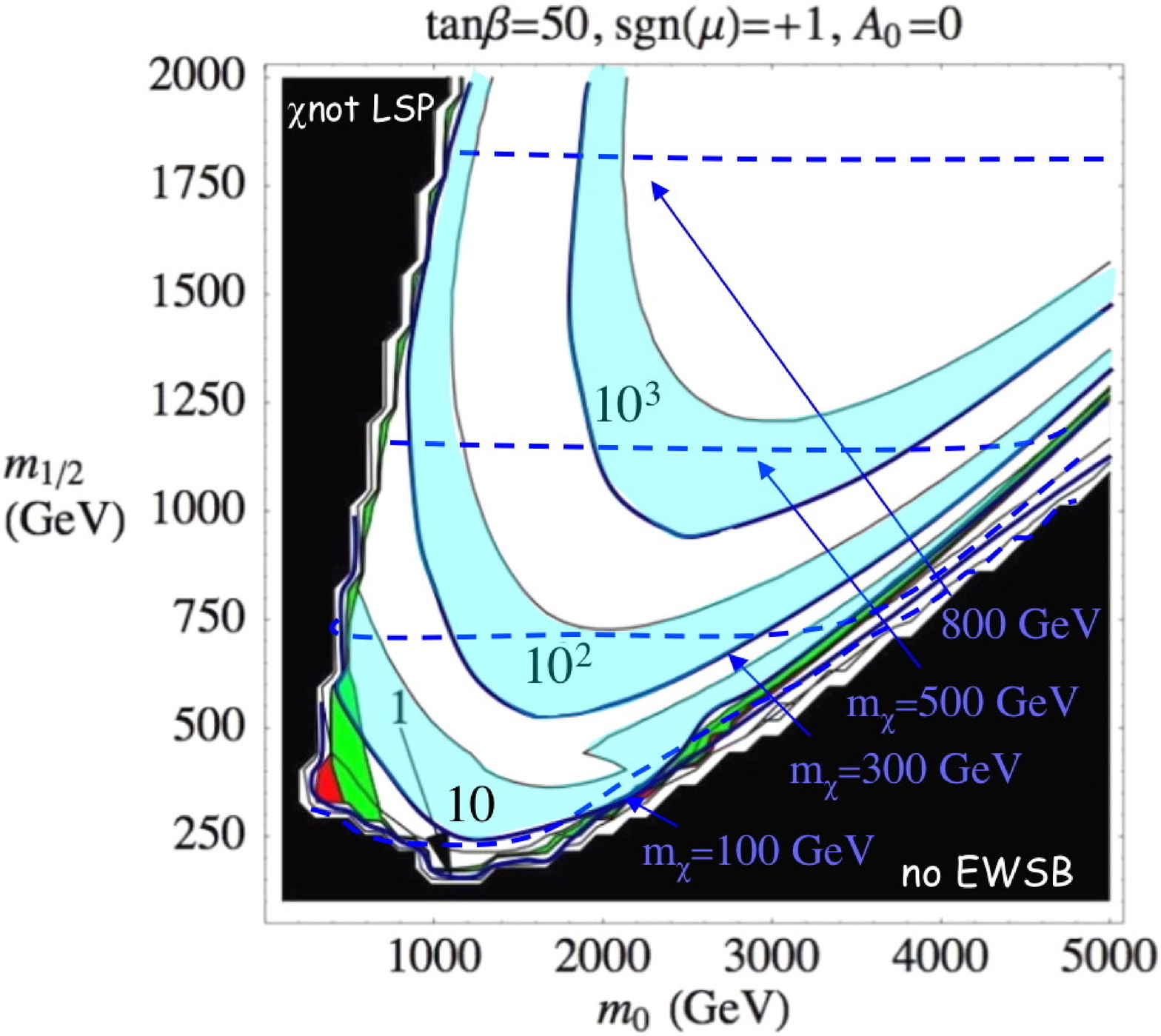}
\includegraphics[scale=0.25]{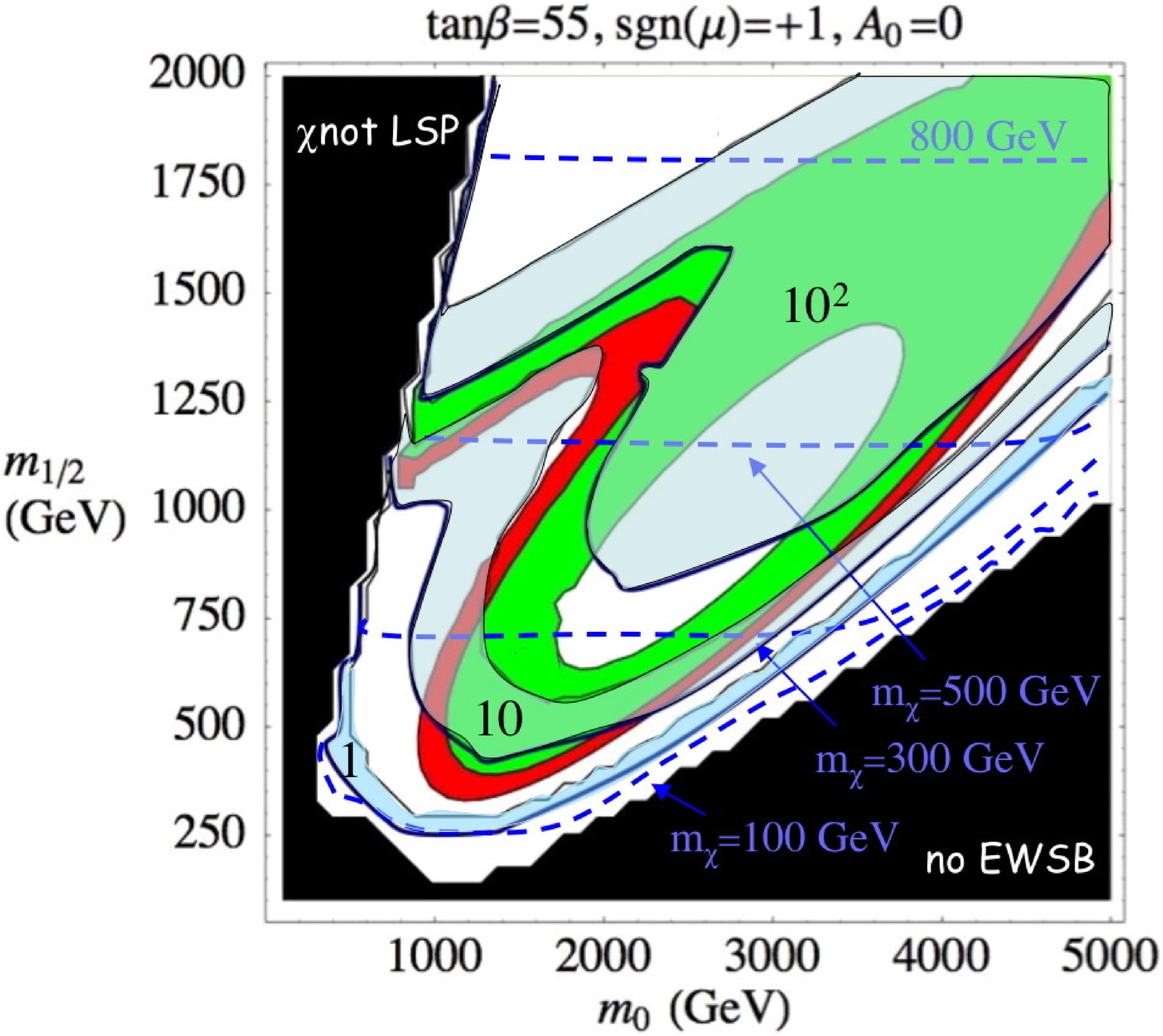}
\caption{Contour plots for the minimum $fd$ needed for a PAMELA disentanglement (upper bounds of the translucent bands) and for the maximum $fd$ allowed by current experimental data (lower bounds of the translucent bands). In the upper panel $\tan \beta = 50$ while in the lower panel $\tan \beta = 55$. The other parameters are $A_{0} = 0$ and $sgn(\mu)$. Black color represents the regions in the parameter space that are excluded either by accelerator bounds or because electroweak symmetry breaking is not achieved or because the neutralino is not the lightest supersymmetric particle. The translucent regions denote the parameter space domains that correspond to models that satisfy both the conditions (\ref{EQc1}) and (\ref{EQc2}), but now keeping fixed values of $fd$ (those indicated inside the regions). Red (dark shaded) are domains with $\Omega h^{2}$ in the WMAP region $0.09 < \Omega h^{2} < 0.13$, while green (light shaded) are the parameter space domains with $0.13 < \Omega h^{2} < 0.3$. Under the line with $fd=1$ there are no models that satisfy both the conditions (\ref{EQc1}) and (\ref{EQc2}). We also show the equi-neutralino mass contours (blue dashed lines).}
\label{FIG50_55_complete}
\end{figure}

\begin{figure}
\centering
\includegraphics[width=6.4cm]{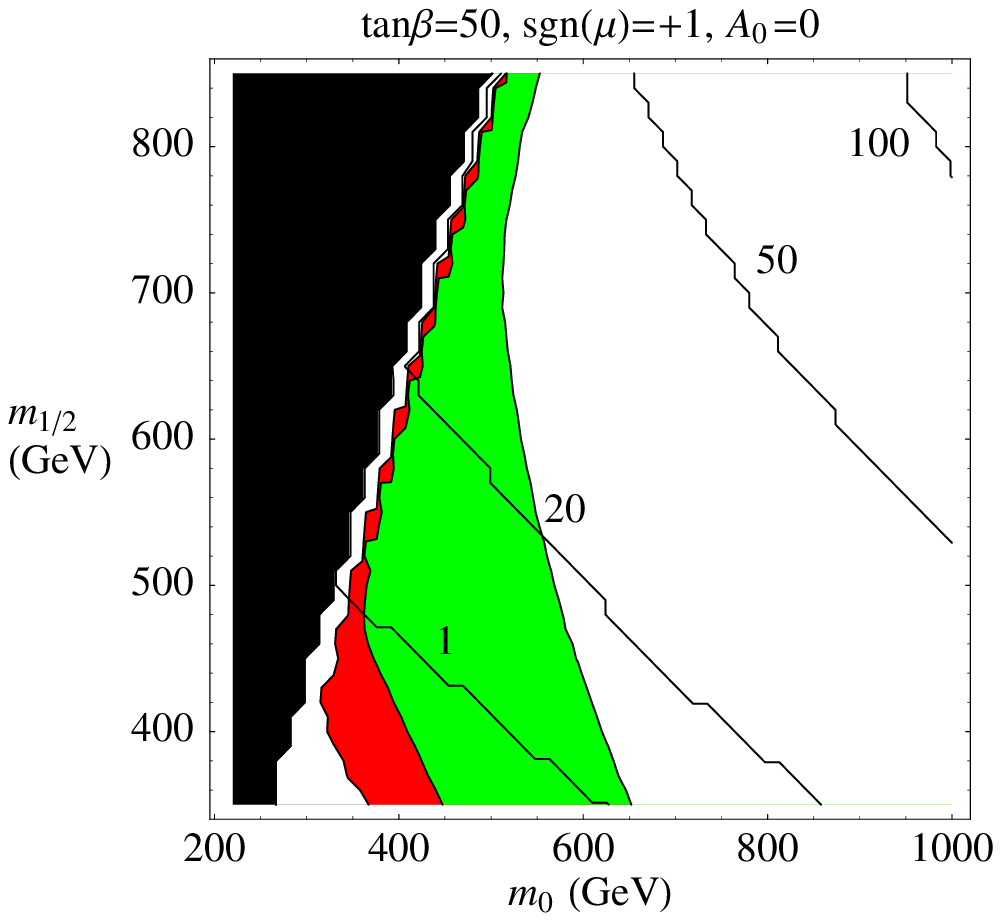}
\includegraphics[width=6.4cm]{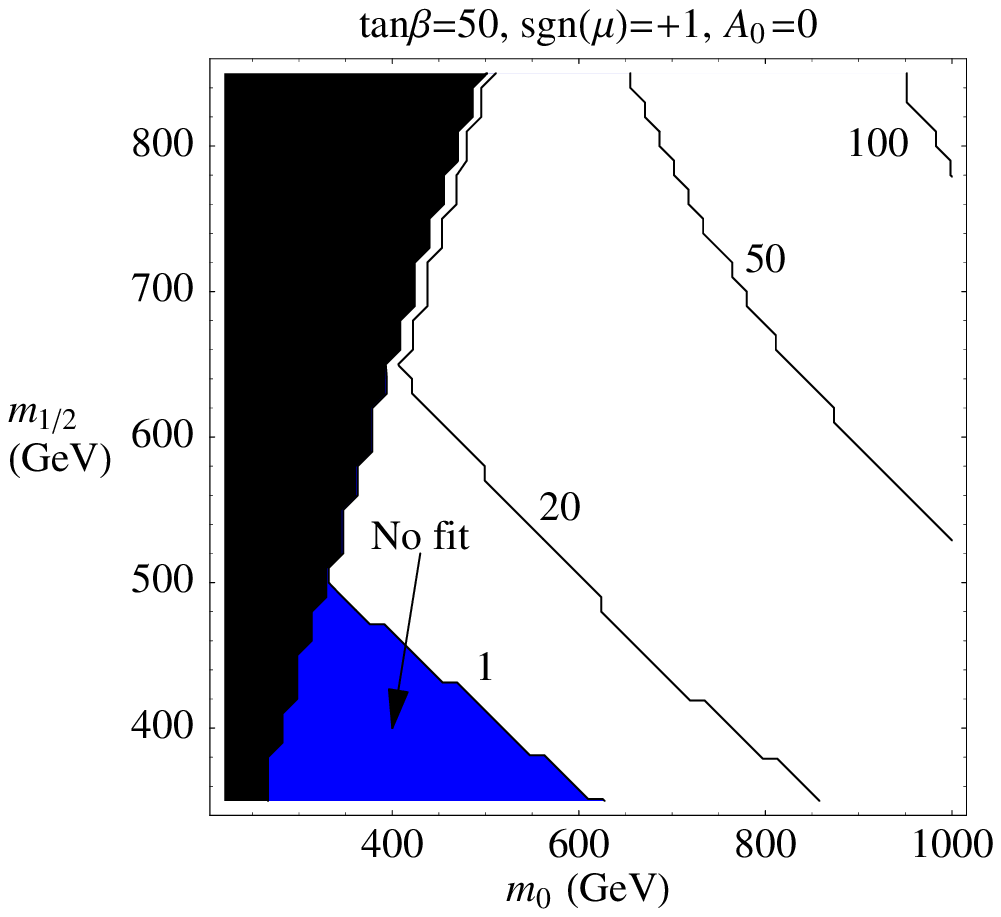}
\caption{Left: equi-clumpiness factors lines as functions of $m_{0}$ and $m_{1/2}$ parameters in the first interesting part of the parameter space (''zoom'' of the WMAP allowed zone) for $\tan \beta=50$. Regions in the parameter space that are excluded either by accelerator bounds or because electroweak symmetry breaking is not achieved or because the neutralino is not the lightest supersymmetric particle are black. Domains with $\Omega h^{2}$ in the WMAP region $0.09 < \Omega h^{2} < 0.13$ are red, while green indicates regions with $0.13 < \Omega h^{2} < 0.3$. Right: the same plot as in the left panel, but with blue region under the line with $fd=1$ in which there are no models that satisfy the both conditions (\ref{EQc1}) and (\ref{EQc2}) simultaneously.}
\label{FIG501fd}
\end{figure}

%==========================================================
For each model we found the minimal value of the clumpiness factor $fd$ needed to satisfy both conditions. As the clumpiness factor is a function of $m_{0}$ and $m_{1/2}$ parameters we made contour plots calculating equi-clumpiness factors lines. We also found the maximally allowed values for $fd$ in order not to violate the condition~(\ref{EQc1}). First, we found the results for different values of $\tan \beta$, presented in figure~\ref{FIG50_55_complete} and in figure \ref{FIG60fd}. Then we examined in more details some interesting zones in those plots.
 The regions in parameter space that are excluded either by accelerator bounds or because electroweak symmetry breaking is not achieved or because the neutralino is not the lightest supersymmetric particle are represented  with black color. Red (dark shaded) color indicates the ($m_{0}$, $m_{1/2}$) domains with $\Omega h^{2}$ in the WMAP \cite{WMAP} region $0.09 < \Omega h^{2} < 0.13$. Green (light shaded) color indicates the parameter space regions with values of $0.13 < \Omega h^{2} < 0.3$. Equi-clumpiness factors lines for the minimal $fd$ needed to disentangle a supersymmetric signal with PAMELA are given with black solid lines.
Beside every line it is indicated the value of the clumpiness factor. Under the line $fd=1$ there are no models that satisfy at the same time the both conditions from equations (\ref{EQc1}) and (\ref{EQc2}).
As an example  we give in the right panel of figure \ref{FIG501fd} a plot with that region colored in blue.
In figures~\ref{FIG50_55_complete} the equi-clumpiness factors lines for the maximal allowed (by the present experimental data) $fd$ are represented by the lower bound of the translucent light blue (very light shaded) regions. In these figures the minimal $fd$ are instead the upper bounds of the same regions. The translucent regions denote the parameter space domains that correspond to models that satisfy both the conditions (\ref{EQc1}) and (\ref{EQc2}), but now keeping fixed values of $fd$ (those indicated inside the regions).

It can be seen that relatively small clumpiness factors (of order 10) are sufficient for PAMELA detection. This is very important, because, even if we consider a DC model as the background flux (that alone already gives a good fit of the experimental data) it is still possible to disentangle a supersymmetric component in a wide region of the parameter space (in comparison with the WMAP allowed zone). The two favourite cases are: $\tan \beta = 55$ (see figure~\ref{FIG50_55_complete}) and $\tan \beta = 60$ (see figure~\ref{FIG60fd}).

\begin{figure}
\centering
\includegraphics[width=7.4cm]{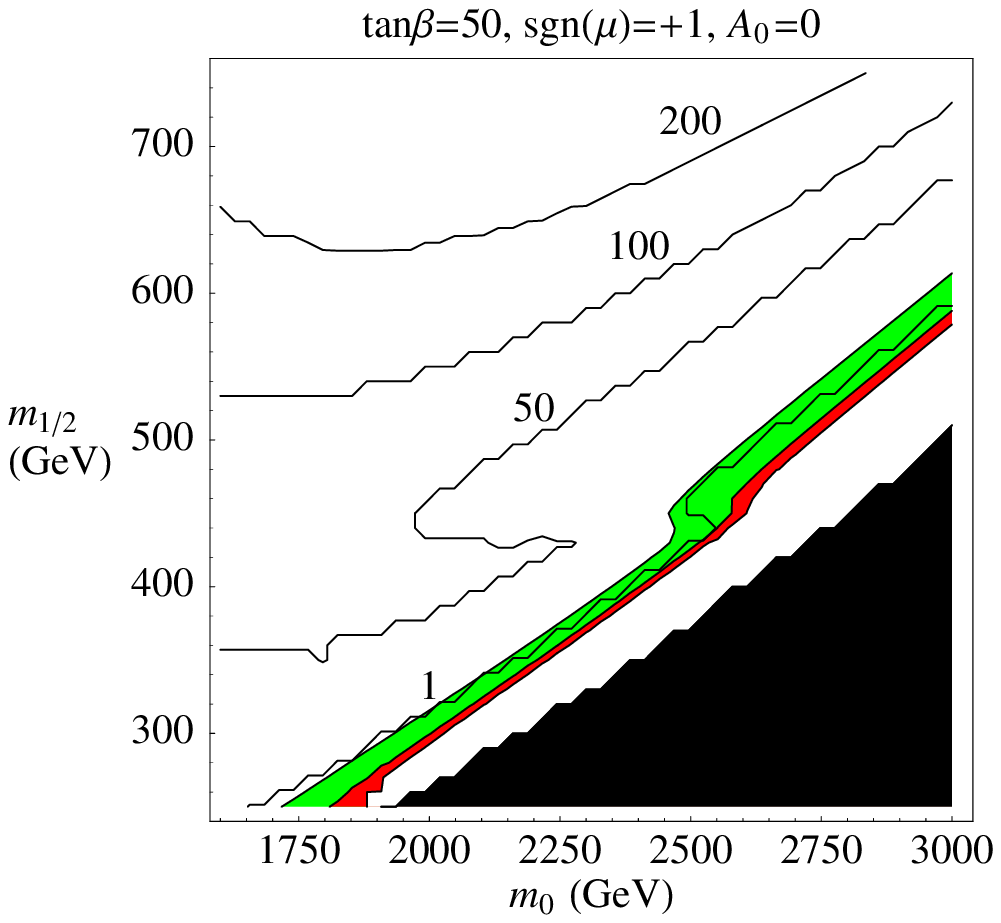}
\caption{Equi-clumpiness factors lines as functions of the $m_{0}$ and $m_{1/2}$ parameters in the second interesting part of the parameter space (''zoom'' of the focus point region) for $\tan \beta=50$. Regions in the parameter space that are excluded either by accelerator bounds or because electroweak symmetry breaking is not achieved or because the neutralino is not the lightest supersymmetric particle are black. Domains with $\Omega h^{2}$ in the WMAP region $0.09 < \Omega h^{2} < 0.13$ are red, while green indicates regions with $0.13 < \Omega h^{2} < 0.3$.}
\label{FIG502fd}
\end{figure}

In figure~\ref{FIG501fd} we presented one of the two particularly interesting regions, namely those inside the cosmologically allowed zone, found for the $\tan \beta=50$ case, i.e. that from the upper panel of figure~\ref{FIG50_55_complete}. The other interesting region, corresponding to the so called focus point region~\cite{focus} in which the neutralino has a significant higgsino component, is presented in figure~\ref{FIG502fd}. We can see that for those cases the region in the parameter space that corresponds to both conditions satisfied, (\ref{EQc1}) and (\ref{EQc2}), and that gives the right relic density $\Omega h^2$, is not so much extended with respect to the total cosmologically allowed zone. The red zone mainly corresponds to small clumpiness factors, i.e. less than 20.


In the $\tan \beta = 55$ case (see the lower panel of figure~\ref{FIG50_55_complete}) there is the maximally extended cosmologically allowed region, and a huge part of it lies between the equi-clumpiness factors lines $fd=1$ and $fd=10$. So, the disentanglement of the supersymmetric signal is achieved without increasing the minimal clumpiness factors.

\begin{figure}
\centering
\includegraphics[width=6.3cm]{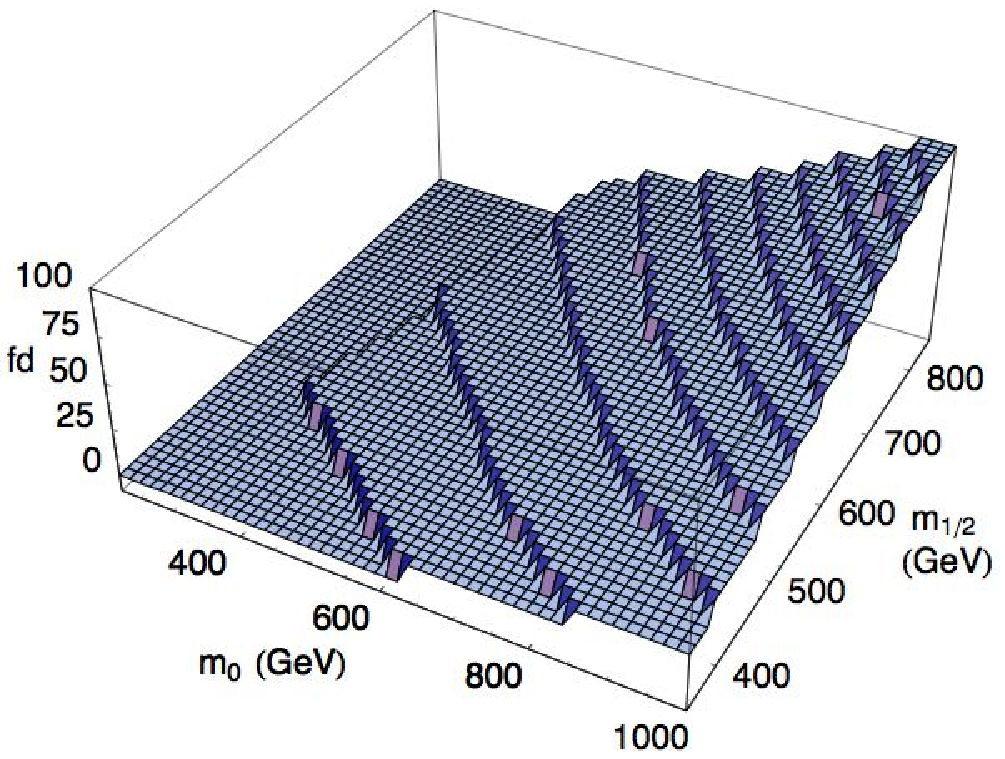}
\includegraphics[width=6.5cm]{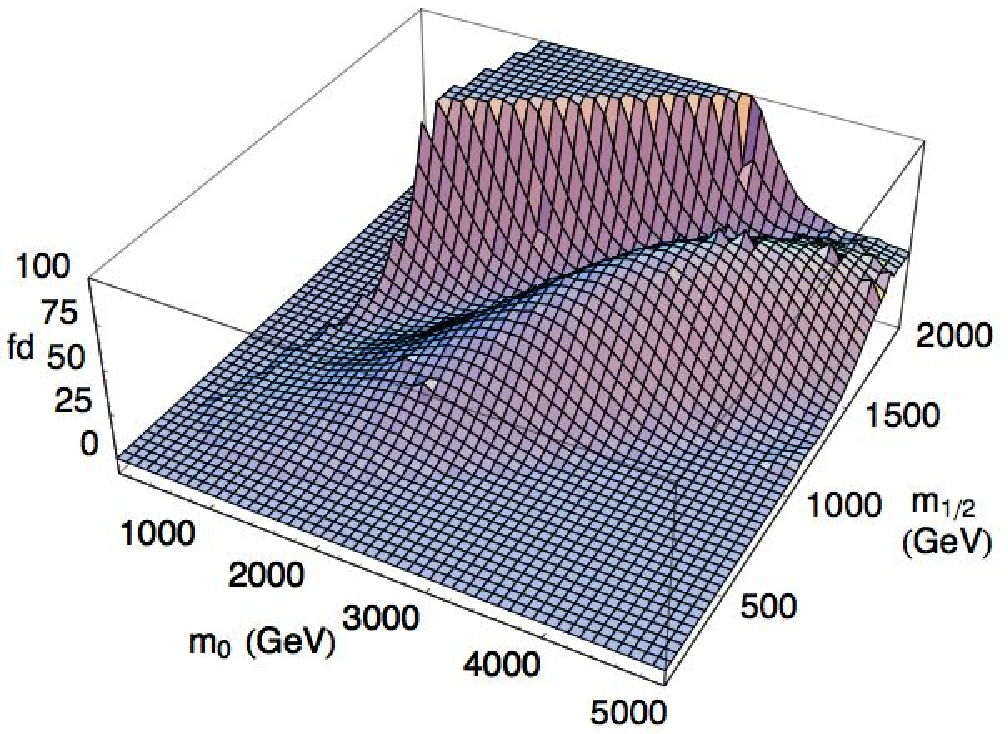}
\caption{Clumpiness factor as a function on the ($m_{0}$, $m_{1/2}$) plane for fixed $\tan \beta = 50$ in the ''zoomed'' zone from the figure \ref{FIG501fd} (left panel) and for $\tan \beta = 55$ for all the masses we considered, i.e.\ the whole mass zone from the lower panel from the figure \ref{FIG50_55_complete} (right panel).}
\label{FIG55fd3d}
\end{figure}

\begin{figure}
\centering
\includegraphics[scale=0.25]{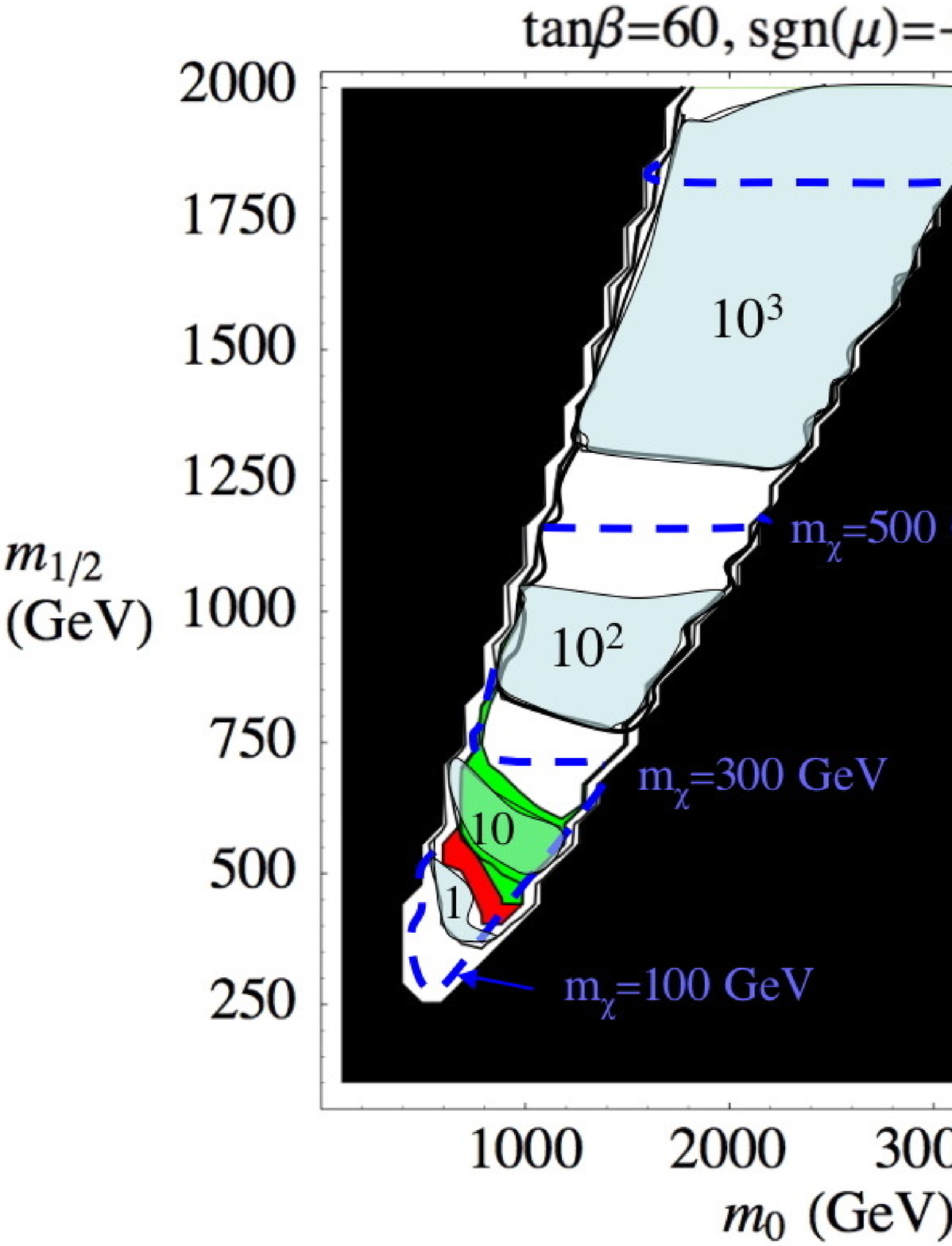}
\caption{Same contour plot as in figure~\ref{FIG50_55_complete} for $\tan \beta= 60$, $A_{0} = 0$ and $sign(\mu)=+1$.}
\label{FIG60fd}
\end{figure}

In figure~\ref{FIG55fd3d} we gave two 3D plots that represent the functional dependance of the minimal $fd$ factor from $m_{0}$ and $m_{1/2}$: the left panel corresponds to figure~\ref{FIG501fd} while the right one corresponds to the lower panel of the figure~\ref{FIG50_55_complete}.

The last case is that of $\tan \beta = 60$ (see figure \ref{FIG60fd}) for which the cosmologically allowed zone (red zone) is completely between the lines with $fd=1$ and $fd=10$.

%==========================================================
\subsection{Some Examples of the Primary Component of the Antiproton Flux}
%==========================================================
In this section we present different contributions to the antiproton flux induced by neutralino annihilations. Fluxes are calculated for different neutralino masses obtained from particular choices of the five mSUGRA parameters that satisfy the WMAP limits on the cosmological relic density and for different choices of the clumpiness factors $fd$.

Higher neutralino masses produce primary contributions that improve high energy data fits but these models need higher clumpiness factors, due to the dependence from the inverse neutralino mass squared $m_\chi^{-2}$ in the primary antiproton flux formula (\ref{eq:pbarsusyflux}).
\begin{figure}
\centering
\includegraphics[width=6.4cm]{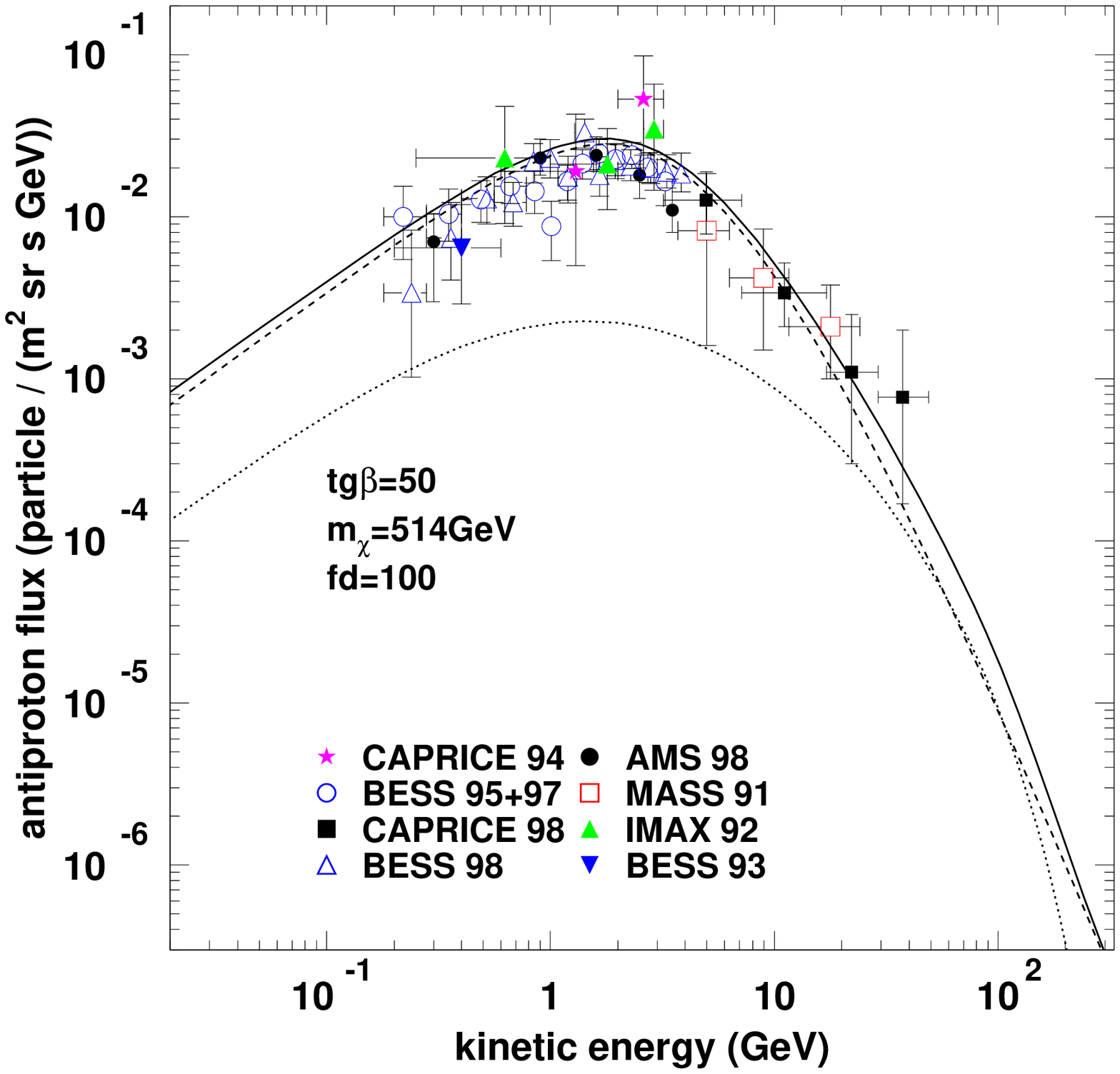}
\includegraphics[width=6.4cm]{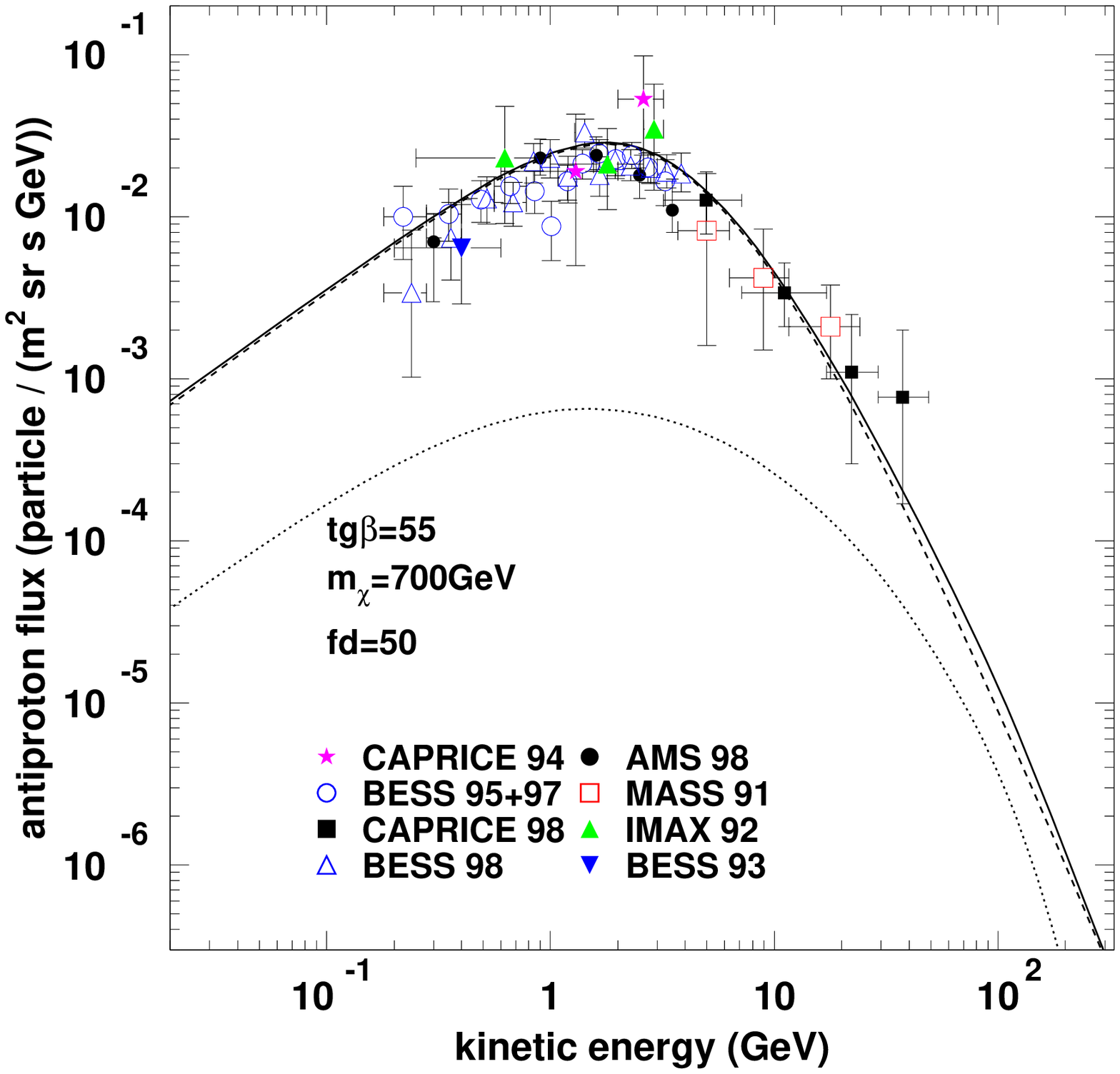}
\includegraphics[width=10cm]{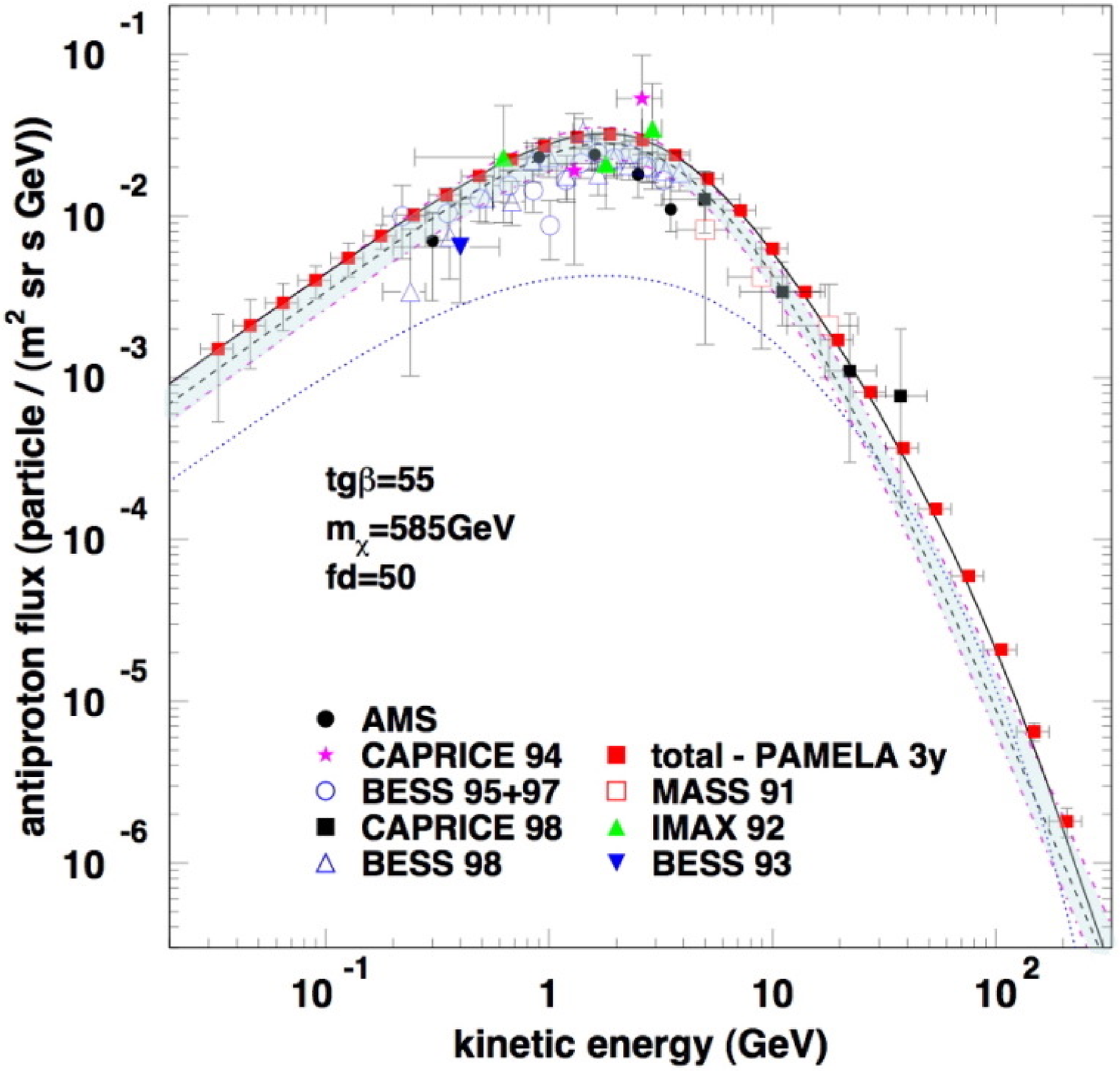}
\caption{Different neutralino annihilations induced contributions to the total antiproton flux with the DC model background that corresponds to the propagation parameters of the best fit of the B/C data. The dashed lines correspond to the background contributions, the punctuated lines correspond to the neutralino induced contributions, while the solid lines correspond to total antiproton fluxes. As an example, in the lower panel we also show the PAMELA expectations for the total flux together with the uncertainty band for the background (translucent band between dot-dashed lines). All the three models are detectable by PAMELA.}
\label{FIG4models}
\end{figure}

Small neutralino masses produce relatively high fluxes at small energies where the data are already fitted with a background contribution only. As a consequence a relatively small part of the mSUGRA parameter space gives the primary contributions that satisfies the experimental data fit.
As can be seen in figure~\ref{FIG4models} there are models that give a total flux that is above the background uncertainty band for $fd\sim 50$ in  the high energy range. 
We can consider two limiting classes of background models. In the first, most favorable class there are models near the lower limit of the uncertainty band. In this case it is always possible to add a supersymmetric contribution that gives fluxes compatible with the experimental data and detectable by PAMELA with a higher clumpiness factor. 
In the second, unfavourite class there are models near the upper bound of the uncertainty band. These models require much higher clumpiness factors and neutralino masses and this reduce the allowed zone in the supersymmetric parameter space.

%--------------------------------------------
\section{Conclusions}
%--------------------------------------------

In the first part of this work we deduced systematically the uncertainties of the most important cosmic rays spectra, with an emphasis on positrons and antiprotons. 

The DR and DRB models systematically overestimate the positron,  protons and helium production.
Electrons are overproduced at low energies, especially  in the case with  the break.
The antiprotons are instead underproduced.
For the model with diffusion and convection  the results are in better agreement with the data.
Further measurements of antiproton and positron spectra, primary to secondary CR ratios and solar modulation, as well as the precise determination and parametrization of important nuclear cross sections~\cite{Kamae:2004xx} seem to be crucial to determine the correct propagation model.  

In the DR and DRB models the antiproton experimental data fits can be easily improved adding different primary components coming from neutralino annihilations or from some other exotic contributions.
In the framework of the DC model, exotic contributions remain possible at high energies ($E > 20$ GeV)
and are not excluded at lower energies, due to the relatively large background uncertainties.

In the second part of this work, we treated the detection of cosmic rays with the upcoming PAMELA experiment.
PAMELA will measure with high statistics and in a wide energy range the spectra of protons, helium and light nuclei~\cite{AP2} and this will help to reduce the uncertainties in the antiproton and positron background fluxes. 
We computed the statistical errors for positrons and antiprotons (see figures figure \ref{FIGe+pamDC} and figure \ref{FIGpamDCpbar}) and we presented in figures~\ref{FIGpbarratio} and~\ref{FIGe+ratio} PAMELA expectations for the antiproton proton ratio and positron charge fraction together with their respective propagation uncertainties. 

Then we studied the possibility to detect an eventual neutralino induced component in the antiproton spectra in the mSUGRA framework assuming the standard component to be the best fit for the DC model. In this case PAMELA will be able to disentangle a neutralino induced component for halo models that has  $fd $ as low as $\sim 1$. 
It is also possible to find models that give a total flux that is above the background uncertainty band but for larger value of $fd $ (see lower panel of figure~\ref{FIG4models}).

Background models near the lower limit of the uncertainty band still give fluxes compatible with the experimental data and detectable by PAMELA but for higher clumpiness factors.

For background models near the upper bound much higher clumpiness factors and neutralino mass  will be needed and this would reduce the allowed zone in the supersymmetric parameter space.

\section*{Acknowledgments}
We would like to thank Igor Moskalenko for the help with the Galprop code, Piero Ullio for the help with DarkSusy,  Elena Vannuccini  for the help with the balloon data.

\end{document}